\documentclass[acmsmall,screen]{acmart}
\usepackage{cite}
\usepackage{xspace}

\usepackage{amsmath,amssymb,amsfonts}
\usepackage{algorithm}
\usepackage{algorithmicx}
\usepackage[noend]{algpseudocode}
\usepackage{enumitem}
\usepackage{hyperref}
\usepackage{graphicx}
\usepackage{textcomp}
\usepackage[table,xcdraw]{xcolor}
\usepackage{booktabs}
\usepackage{tcolorbox}
\usepackage{multirow}
\usepackage{listings}
\usepackage{caption}
\usepackage[normalem]{ulem}
\usepackage{bbding}

\usepackage{wrapfig}
\usepackage{setspace}

\useunder{\uline}{\ul}{}
\AtBeginDocument{%
  }

\algnewcommand{\algorithmiccontinue}{\textbf{continue}}
\algnewcommand{\Continue}{\State \algorithmiccontinue}

\newcommand{\tool}{SR-Eval\xspace}

\setcopyright{acmlicensed}
\copyrightyear{2025}
\acmYear{2025}
\acmDOI{XXXXXXX.XXXXXXX}
\acmISBN{978-1-4503-XXXX-X/2018/06}

\begin{document}

\title{SR-Eval: Evaluating LLMs on Code Generation under Stepwise Requirement Refinement\\
}


\author{Zexun Zhan}
\affiliation{%
  \institution{Sichuan University}
  \city{Chengdu}
  \country{China}}
\email{zhanzexun@stu.scu.edu.cn}

\author{Shuzheng Gao}
\affiliation{%
  \institution{The Chinese University of Hong Kong}
  \city{Hong Kong}
  \country{China}
}
\email{szgao23@cse.cuhk.edu.hk}

\author{Ruida Hu}
\affiliation{%
 \institution{Harbin Institute of Technology}
 \city{Shenzhen}
 \country{China}}
\email{200111107@stu.hit.edu.cn}

\author{Cuiyun Gao}
\affiliation{%
  \institution{Harbin Institute of Technology}
  \city{Shenzhen}
  \country{China}}
\email{gaocuiyun@hit.edu.cn}






\begin{abstract}
Large language models (LLMs) have achieved remarkable progress in code generation. However, existing benchmarks mainly formalize the task as a static, single-turn problem, overlooking the stepwise requirement changes and iterative workflows in real-world software development. In practice, developers seldom work from a complete requirement; instead, requirements are incrementally refined through successive iterations of development. This mismatch limits the understanding of how well LLMs can support real-world development workflows. However, constructing benchmarks for such scenarios is challenging, as real-world interaction traces are not publicly available. Additionally, the test cases for different rounds need to be semantically aligned with each requirement and discriminative enough to facilitate accurate evaluation.

To bridge this gap, we present \tool, a benchmark specifically designed to assess LLMs on iterative code generation under \textbf{S}tepwise requirements \textbf{R}efinement. \tool spans both function-level and repository-level tasks in Python and Java, enabling fine-grained and progressive evaluation across evolving requirements. The construction of \tool follows a carefully designed pipeline that first leverages a multi-agent-based requirement generation method to simulate the development process and recover the multi-round interaction process from final requirements, then employs a semantic-aware discriminative test case generation component to ensure discriminative and consistent evaluation at each turn. \tool comprises 443 multi-turn tasks and 1,857 questions at both function and repository levels. Using \tool, we evaluate 11 representative LLMs with three prompting strategies that simulate different usage patterns.
Results show that iterative code generation under stepwise requirement refinement remains highly challenging: the best-performing model achieves only 22.67\% completion rate on function-level tasks and 20.00\% on repository-level tasks. We further observe that prompting strategies substantially influence performance, highlighting the need for the development of advanced methods. 

\end{abstract}




\maketitle

\section{Introduction}\label{sec:intro}
Large language models (LLMs) have been widely applied to software development tasks\citep{DBLP:journals/corr/abs-2107-03374,DBLP:journals/corr/abs-2108-07732,DBLP:journals/corr/abs-2308-12950,DBLP:conf/iclr/NijkampPHTWZSX23,DBLP:conf/icse/GaoMG000L24}, greatly enhancing development efficiency and powering AI-assisted tools like GitHub Copilot\citep{github_copilot} and Cursor\citep{Cursor}. These tools support developers in a wide range of coding scenarios, from function-level coding to repository-scale development involving complex code contexts. As LLMs continue to evolve, understanding their capabilities and limitations in realistic programming workflows has become increasingly critical.

To evaluate LLM capabilities in code generation, existing code generation benchmarks have substantially advanced evaluation in this space. For instance, HumanEval~\citep{DBLP:journals/corr/abs-2107-03374} and MBPP~\citep{DBLP:journals/corr/abs-2108-07732} focus on function-level program synthesis with well-defined specifications. Recently advanced datasets like LiveCodeBench~\citep{DBLP:conf/iclr/JainHGLYZWSSS25} and DevEval~\citep{DBLP:conf/acl/Li0ZLLZWLFWDZZD24} extend to algorithmic problems~\citep{hendrycks2021measuring,DBLP:conf/iclr/JainHGLYZWSSS25,li2022competition} and even repository-level tasks \citep{DBLP:conf/acl/Li0ZLLZWLFWDZZD24,DBLP:conf/iclr/YangJZLYWPMSNY025,feng2024complexcodeeval,yu2024codereval}.

Despite these advances, current evaluation paradigms have a critical limitation. Existing benchmarks treat code generation as a single-pass process with complete and static requirements, where models receive a single detailed prompt and produce a finished solution without any need for further refinement. This process emulates the idealized waterfall-like development methodology~\citep{DBLP:conf/icse/Royce87} that misaligns with the incremental, conversational workflows in modern agile development practices~\citep{manifesto2001manifesto}. In practice, requirements are inherently dynamic and rarely complete at the outset~\citep{10.1145/800027.808439,wiegers2013software,DBLP:journals/pacmse/Mu00YZWL024}. Modern agile software development process begins with high-level and ambiguous descriptions that are progressively refined throughout development. 
Developers follow an iterative development methodology involving multi-turn interactions, wherein requirements and implementation undergo co-evolution through repeated cycles of requirement refinement and corresponding implementation adaptation~\citep{swartout1982inevitable,basil2012iterative}.
Therefore, there is a pressing need for benchmarking LLM's capacity to handle iterative code development processes under 
stepwise requirement refinement.

Constructing a benchmark that reflects the iterative nature of software development presents several challenges. First, it is difficult to collect real-world developer–LLM dialogues. These iterative conversations between developers and AI assistants typically unfold within private development environments or proprietary corporate platforms, making these interaction data unavailable~\citep{chen2025empirical,DBLP:journals/corr/abs-2503-22625}.  Moreover, public code repositories typically contain only the refined requirements and code that have undergone multiple rounds of iteration, without preserving the intermediate development processes that led to the final version.
Second, it is challenging to create high-quality test cases for each iteration of requirement updates. The test cases should be semantically-aligned and discriminative enough to validate whether the code satisfies the specific incremental requirement of each turn, rather than merely validating the broader functionality established in previous iterations. 
Although some recent work~\citep{DBLP:conf/iclr/HanHSH25, DBLP:journals/corr/abs-2503-22688} also focused on multi-turn code generation, they only focus on feedback-driven bug fixing and instruction-following scenarios, which differ from the common usage scenario of LLMs in IDEs or chat environments~\citep{Cursor,github_copilot} that involves proactively evolving requirements.

\textbf{Benchmark \tool.} In this paper, 
we introduce \tool, a new benchmark explicitly designed to evaluate LLMs for iterative code generation under \textbf{S}tepwise requirement \textbf{R}efinement rather than static one-shot complete requirement. 
\tool spans both function-level and repository-level programming tasks, covering Python and Java. In total, it includes 314 function-level tasks with 1,348 instruction turns, and 129 repository-level tasks with 509 instruction turns. 
The construction of \tool follows a rigorous pipeline. 
Starting from a curated seed dataset, we first leverage 
a multi-agent–based requirement generation process that simulates real-world iterative development by transforming complete requirement into multi-round, evolving instructions. Then, the semantic-aware discriminative test case generation method is proposed to synthesize discriminative test cases for each turn, ensuring both correctness validation and precise alignment with the specific incremental requirements of each turn.

\textbf{Main findings and implications.} Based on \tool, we evaluate 11 representative LLMs that span a diverse range of characteristics~\citep{yang2025qwen3,DBLP:journals/corr/abs-2501-12948,liu2024deepseek,gpt-5,openai2025gptoss120bgptoss20bmodel}. The selected models comprise both commercial and open-source models 
with varying parameter scales and different levels of reasoning capabilities.
We further investigate the impact of different prompting strategies on iterative code generation.
Based on extensive experiments, our study reveals several key findings:
\begin{enumerate}
    \item {The performance of state-of-the-art LLMs remains limited under iterative development scenarios}. 
    \item Reasoning capabilities alone do not guarantee improvements, while prompting strategies are critical for balancing accuracy and efficiency.
    \item {{Our proposed multi-agent–based requirement generation and semantic-aware discriminative test case generation could create high-quality evaluation instructions and improve test cases in detecting whether the code meets the requirements of each round.}}
\end{enumerate}

Our contributions are summarized as follows:
\begin{itemize}
    \item We introduce \tool, the first 
    benchmark for evaluating 
    iterative code generation {under stepwise requirement refinement scenarios rather than static and complete requirement}
    {that encompass both} function-level and repository-level tasks.
    \item {We propose a data construction framework that contains multi-agent-based requirement generation and semantic-aware discriminative test case generation to produce high-quality instructions and test cases.} 
    \item We conduct a systematic evaluation across a diverse range of models and {provide} 
    insights into their strengths and weaknesses under iterative development scenarios.
\end{itemize}

\section{Related Work}\label{sec:related}

\subsection{LLMs for Code}\label{sec:related_llm4code}

Large language models (LLMs) have recently achieved remarkable progress across various code-related tasks, including code generation, code editing, program repair, and test case generation \citep{DBLP:journals/corr/abs-2107-03374,DBLP:conf/kbse/GaoWGWZL23,gao2024search,DBLP:conf/iclr/YangJZLYWPMSNY025,DBLP:journals/pacmse/Yuan0DW00L24}, substantially advancing automated software development.
Early efforts focused on pretraining LLMs on large-scale code corpora, resulting in foundational models such as CodeX \citep{codex}. Building on this foundation, a number of code-specialized models have emerged, such as CodeLlama \citep{DBLP:journals/corr/abs-2308-12950}, StarCoder \citep{li2023starcoder}, Qwen-Coder \citep{hui2024qwen2} and DeepSeekCoder \citep{guo2024deepseek}.
In parallel, general-purpose LLMs like GPT-5 \citep{gpt-5}, DeepSeek-V3 \citep{liu2024deepseek} and Qwen3 variants \citep{yang2025qwen3} have also shown strong performance on these
tasks.

More recently, the focus has shifted from scaling models during pretraining to enhancing capabilities at test time, known as test-time scaling (TTS) \citep{snell2024scaling,zhang2025survey}. Studies show that TTS techniques can improve LLMs’ reasoning and problem-solving abilities in complex domains such as programming and mathematics \citep{DBLP:journals/corr/abs-2501-12948,learning-to-reason-with-llms}. Notable examples include OpenAI’s o1 and o3 \citep{learning-to-reason-with-llms}, DeepSeek-R1 \citep{DBLP:journals/corr/abs-2501-12948}, and Alibaba’s QwQ \citep{qwq32b}. Some models, such as gpt-oss series \citep{openai2025gptoss120bgptoss20bmodel} and Qwen3 series \citep{yang2025qwen3}, further support dynamic switching between reasoning and standard generation modes, or adjusting the reasoning level, enabling users to balance performance and efficiency.


\subsection{Code LLMs Benchmarks}\label{sec:related_bench}


Early representative benchmarks such as HumanEval \citep{DBLP:journals/corr/abs-2107-03374} and MBPP \citep{DBLP:journals/corr/abs-2108-07732} primarily focus on well-structured, small-scale, function-level algorithmic problems. These benchmarks are designed to assess a model’s basic code generation and comprehension abilities. To introduce greater challenge and realism, benchmarks like APPS \citep{hendrycks2021measuring}, CodeContests \citep{li2022competition}, and LiveCodeBench \citep{DBLP:conf/iclr/JainHGLYZWSSS25} incorporate competitive programming problems, better reflecting real-world requirements for logical reasoning and problem-solving. ClassEval \citep{du2023classeval} further extends the task granularity to the class level, enhancing structural modeling complexity.
In terms of natural language to code generation, BigCodeBench \citep{DBLP:conf/iclr/ZhuoVCH0WYZHPB025} and AutoCodeBench \citep{chou2025autocodebenchlargelanguagemodels} evaluated a model’s ability to follow complex instructions and use external libraries. 
%
Recent work has also moved beyond function-level tasks to repository-level evaluation, aligning more closely with real-world software development workflows. Benchmarks such as CrossCodeEval \citep{ding2023crosscodeeval}, DevEval \citep{DBLP:conf/acl/Li0ZLLZWLFWDZZD24}, CoderEval \citep{yu2024codereval}, and ComplexCodeEval \citep{feng2024complexcodeeval} introduce repository-level tasks to examine a model’s ability to understand context dependencies, inter-module relationships, and multi-file project structures. 
{InterCode \citep{yang2023intercode} and ConvCodeWorld \citep{DBLP:conf/iclr/HanHSH25} introduce multi-turn interaction mechanisms but only focus on feedback-driven bug fixing.}
CodeFlowBench \citep{DBLP:journals/corr/abs-2504-21751} focuses on the ability to reuse existing functions in competitive programming.
{CodeIFBench \citep{DBLP:journals/corr/abs-2503-22688} is the most related work to us. However, it primarily targets a narrow set of nine synthetically generated instruction-following tasks, limiting the diversity of instructions and failing to reflect the complexity of real-world instruction evolution.}
\section{\tool}\label{sec:benchmark}

\begin{figure*}[t]
  \centering
  \includegraphics[width=\textwidth]{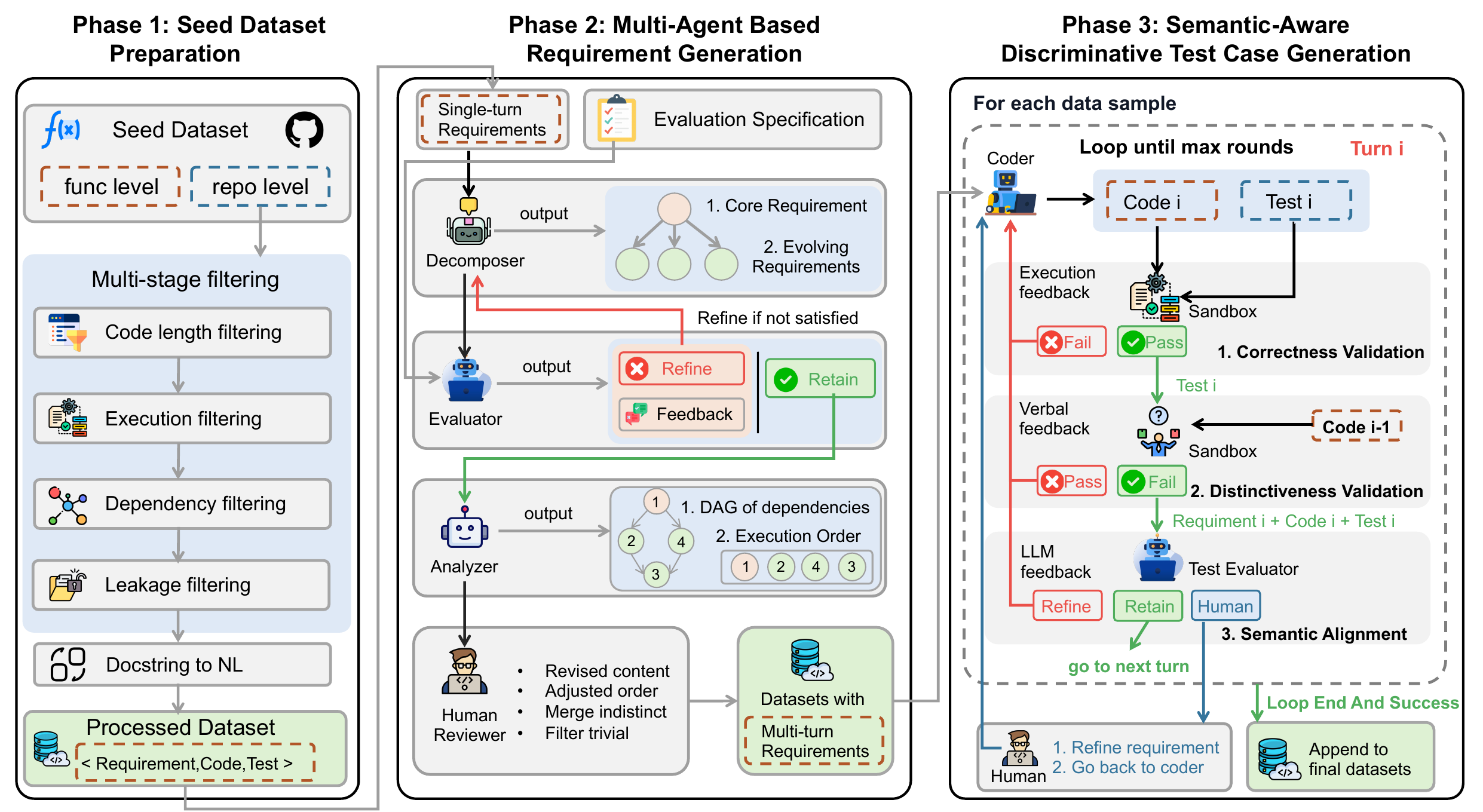}
  \caption{The overview of our benchmark construction pipeline.}
  \label{fig:overview}
\end{figure*}

In this section, we describe the benchmark construction pipeline as shown in Fig.~\ref{fig:overview}, which comprises three parts: seed dataset preparation, multi-agent–based requirement generation, and semantic-aware discriminative test case generation.


\subsection{Benchmark Construction Pipeline}\label{sec:pipeline}

\subsubsection{Seed Dataset Preparation}\label{sec:seed_prepare}

To construct \tool, we recognize that multi-turn conversations between developers and AI programming assistants are typically confined to private or proprietary platforms. Accordingly, we build upon existing high-quality single-turn code generation datasets and extend them to multi-turn iterative development scenarios.


\paragraph{Data Collection}

For the \textbf{function-level} tasks, we collected 148 Python tasks from
BigCodeBench-Hard~\citep{DBLP:conf/iclr/ZhuoVCH0WYZHPB025} and 188 Java tasks from AutoCodeBench~\citep{chou2025autocodebenchlargelanguagemodels}. These datasets are well-suited
for simulating iterative development since their complex task descriptions involve multiple function calls and nested logic.
Crucially, both datasets mitigate data contamination. BigCodeBench-Hard is carefully curated with extensive human annotations, and its low pass@1 score (35.8) suggests resistance to
model overfitting. Moreover, AutoCodeBench was released after the training cutoff of the models we evaluate.

For the \textbf{repository-level} tasks, we collected 1,825 Python samples from DevEval~\citep{DBLP:conf/acl/Li0ZLLZWLFWDZZD24} and 106 Java samples from MRGBench~\citep{li2025mrg}. Both datasets provide function-completion tasks within a repository context. The substantial length of the target functions makes them well-suited for iterative  development scenarios. Sourced from real-world GitHub projects, these tasks ensure the practical relevance of our evaluation.

\paragraph{Data Preprocessing for Repository-level Tasks}

To construct a rigorous benchmark for repository-level tasks, we apply a multi-stage filtering pipeline. First, we remove samples that fail to pass the provided test cases in their original repository environments. Second, we discard samples whose function bodies contained fewer than 10 lines of code, since such trivial cases tend to resemble single-shot code completion tasks rather than incremental development scenarios. Third, for DevEval, we exclude samples with \textit{intra-function} dependencies, which do not require repository-level context, as well as those with \textit{cross-file} dependencies, which introduce confounding variables from retrieval strategies and fall outside our focus on stepwise refinement. After these steps, we retain 140 Python samples and 48 Java samples.

Given that DevEval was released prior to the evaluation of our target models, we further conducted a contamination check. Specifically, we prompted three representative open-source models—DeepSeek-V3.1, Qwen3-Coder-480B-A35B-Instruct, and gpt-oss-120b—to generate implementations using only function signatures and Docstrings as inputs. Since all 140 tasks inherently require repository context, any model-generated code that nevertheless passed the reference test cases was considered as a strong indicator of data leakage. We removed the union of such contaminated samples across the three models, eliminating 24 tasks in total and retaining 116 uncontaminated samples. For MRGBench, we manually collected functions from the latest repository to mitigate data contamination. After these steps, we retain 116 Python samples and 48 Java samples.

Finally, to support the subsequent requirement generation (Section~\ref{sec:decomposition}), we converted function signatures and docstrings into natural language programming instructions using a large language model. The implementation details of this transformation are described in Section~\ref{sec:impl_detail}.

\begin{algorithm}[t]
\caption{Semantic-aware Discriminative Test Case Generation}
\label{alg:triple_verified}
\small
\begin{algorithmic}[1]
\Require Data sample $d$, llm Coder, max iteration $M$
\State Initialize $turn \gets 1$, $iter \gets 1$
\State Initialize $defect\_code, defect\_test, feedback \gets None$
\State Initialize $trajectory \gets [~]$
\While{$turn \leq d.total\_turn$ \textbf{and} $iter \leq M$}
    \State $r \gets d.GetRequirement(turn)$
    \If{d.IsRepositoryLevel}
        \State $(code, test) \gets Coder(turn, r, defect\_code, defect\_test, feedback, d.code, d.test)$
    \Else
        \State $(code, test) \gets Coder(turn, r, defect\_code, defect\_test, feedback)$
    \EndIf
    \State \texttt{// Stage 1: Correctness Validation}
    \If{\textbf{not} UnifiedCheck$(code, test, \texttt{correctness})$}
        \State $(defect\_code, defect\_test, feedback) \gets (code, test, \text{Stage1Feedback})$
        \State $iter \gets iter+1$; \Continue
    \EndIf
    \State \texttt{// Stage 2: Distinctiveness Validation}
    \If{$turn > 1$ \textbf{and not} UnifiedCheck$(trajectory[-1].code, test, \texttt{distinctiveness})$}
        \State $(defect\_code, defect\_test, feedback) \gets (code, test, \text{Stage2Feedback})$
        \State $iter \gets iter+1$; \Continue
    \EndIf
    \State \texttt{// Stage 3: Sematic Alignment}
    \State $(decision, feedback) \gets LLMEvaluator(r, code, test)$
    \If{$decision = retain$}
        \State $trajectory.append((code,test))$
        \State $turn \gets turn+1$; $iter \gets 0$; $feedback \gets None$
    \ElsIf{$decision = refine\_test$}
        \State $(defect\_code, defect\_test) \gets (code, test)$; $iter \gets iter+1$
    \ElsIf{$decision = refine\_requirement$}
        \State human\_interrupt(); $iter \gets iter+1$
    \EndIf
\EndWhile
\If{$|trajectory| = d.total\_turn$}
    \State save\_code\_test($d, trajectory$)
\EndIf
\end{algorithmic}
\end{algorithm}

\subsubsection{Multi-Agent Based Requirement Generation}\label{sec:decomposition}

This part aims to construct realistic evolving requirements that mirror the evolution of software specifications in practice. To this end, we develop a \textbf{multi-agent-based requirement generation module}, which decomposes an initially complex requirement into a core requirement, followed by a sequence of evolving requirements that progressively add new features, address edge cases, or refine the design. This design reflects
the dynamic nature of real-world software development, where requirements continuously evolve and design decisions are iteratively revised.


\paragraph{Agent's Responsibilities}

As illustrated in Figure~\ref{fig:overview}, our framework employs three specialized agents—the \emph{Decomposer}, the \emph{Evaluator}, and the \emph{Analyzer}.

\begin{itemize}
\item \textbf{Decomposer}: Responsible for breaking down a complex requirement into a core functionality and multiple supplementary requirements. It does not assume a fixed execution order.
\item \textbf{Evaluator}: Responsible for assessing the quality of the decomposed requirements and providing actionable feedback for refinement when necessary.
\item \textbf{Analyzer}: Responsible for analyzing the decomposed requirements, constructing a directed acyclic graph (DAG) to represent dependency relationships, and determining the final execution order.
\end{itemize}

\paragraph{Generation Process}

Given an initial complex requirement, the \emph{decomposer} first splits it into a core requirement and a sequence of evolving requirements
that progressively introduce new features, handle edge cases, or refine the design. To maintain tractability, the decomposition is constrained to between two and five turns.

Next, the \emph{Evaluator} assesses the decomposed requirements along four dimensions—\emph{\textbf{Testability}}, \emph{\textbf{Completeness}}, \emph{\textbf{Distinctiveness}}, and \emph{\textbf{Scenario Authenticity}}. Based on the assessment, it either labels the instructions as \textsc{Retain} or requests \textsc{Refine}. In the latter case, it provides specific, actionable feedback to guide the \emph{decomposer} in improving the instructions. In the former, the instructions are passed to the \emph{analyzer}. Details of the evaluation criteria are presented in Section~\ref{sec:eval_spec}.

Finally, the \emph{analyzer} processes
the refined requirements, constructs a directed acyclic graph (DAG) rooted in the core requirement, and applies topological sorting to determine a conflict-free execution order. The entire process is capped at five iterations. Successful cases are retained in the dataset, while those exceeding the iteration limit are manually revised following the same evaluation guidelines. We also excluded a small number of unsuitable cases, such as overly trivial tasks or excessively fragmented decompositions. 
After these steps, we retain 89 and 40 repository-level samples for Python and Java, respectively, and 142 and 172 function-level samples for Python and Java, respectively.
All agent prompts are publicly available in our GitHub repository. 

\paragraph{Evaluation Specification}\label{sec:eval_spec}

To capture the incremental and multi-turn nature of iterative software development, we design our evaluation specification around five dimensions.
{The first four are used by the evaluator agent in Section~\ref{sec:decomposition}, where instruction order has not yet been determined. In contrast, human evaluation incorporates all five dimensions.}

\begin{itemize}
    \item \textbf{Testability}: Each requirement should be formulated in a way that allows verification through explicit unit tests, without relying on vague or subjective descriptions (\textit{e.g.}, “make the code look better” or “improve efficiency”).
    \item \textbf{Completeness}: The final program, after following the full sequence of requirements, should fully satisfy the original high-level task.
    \item \textbf{Distinctiveness}: Each requirement should introduce a unique modification or extension, avoiding redundancy or overlap with other requirements in the sequence.
    \item \textbf{Scenario Authenticity}: The decomposed requirements should reflect realistic software development practices and align with scenarios commonly encountered in real-world projects.
    \item \textbf{Logical Coherence}: The instructions should form a clear and logically consistent development trajectory.
\end{itemize}

\begin{algorithm}[t]
\caption{Unified Check (for Correctness and Distinctiveness)}
\label{alg:unified_check}
\small

\begin{algorithmic}[1]
\Function{UnifiedCheck}{$code, test, mode$}
    \If{IsRepositoryLevel}
        \State backup $\gets$ repo.snapshot()
        \State repo.replace\_function($code$)
        \State repo.insert\_test($test$)
        \State $(status, feedback) \gets exec\_in\_repo(repo)$
        \State repo.restore(backup)
    \Else
        \State $(status, feedback) \gets exec\_in\_sandbox(code, test)$
    \EndIf
    \If{$mode = correctness$}
        \State \Return $(status = pass)$
    \ElsIf{$mode = distinctiveness$}
        \State \Return $(status = fail)$
    \EndIf
\EndFunction
\end{algorithmic}
\end{algorithm}

\subsubsection{Semantic-Aware Discriminative Test Case Generation}\label{sec:codegen}

This module aims to construct both reference implementations and reliable test cases for each turn of requirements. The test cases enable process-oriented evaluation in iterative development, while the reference implementations serve as golden contexts for assessing model performance under the assumption that historical information is accurate.

\paragraph{Challenges of LLM-Driven Test Case Generation}

LLM-based test case generation presents several challenges. First, ensuring code correctness requires dynamic interaction between the model and the execution environment~\citep{DBLP:conf/iclr/ZhuoVCH0WYZHPB025,chou2025autocodebenchlargelanguagemodels}. Second, test cases must be turn-sensitive, capturing the incremental nature of evolving requirements rather than merely validating generic functionality. Third, the phenomenon of implicit requirements in test cases~\citep{chou2025autocodebenchlargelanguagemodels} often introduces ambiguity. For example, test functions may expect natural language outputs for edge cases that are not explicitly specified in the requirement, causing misalignment between generated code and test expectations, which may underestimate model capability.
To address these issues, we propose a \textit{Semantic-Aware Discriminative Test Case Generation} method consisting of three stages: \textit{correctness validation}, \textit{distinctiveness validation}, and \textit{fine-grained semantic alignment}, as illustrated in Figure~\ref{fig:overview}.

\paragraph{Preliminary Generation}
For each multi-turn sample $d$, we follow the procedure outlined in Algorithm~\ref{alg:triple_verified}. Specifically, for each turn, the \textit{Coder} 
generates candidate code and corresponding test cases based on the current instruction and turn index. For function-level datasets, we deliberately exclude the reference implementations and test cases from the seed dataset to avoid trivial copying. In contrast, repository-level tasks often involve contextual APIs. To prevent the model from fabricating non-existent APIs, we provide the reference implementation and tests as grounding context.

\paragraph{Correctness and Distinctiveness Validation}
The generated code and tests then undergo correctness validation. For function-level tasks, we execute both in an isolated sandbox and collect execution results. For repository-level tasks, we first locate the target function, replace it with the generated implementation, inject the generated tests, and run them within the repository environment. If all tests pass, the process proceeds to distinctiveness validation; otherwise, the execution feedback is returned to the \textit{Coder} for refinement.

Distinctiveness validation ensures that tests meaningfully differentiate successive implementations. When $current\_turn > 1$, the tests produced in the current turn are executed against the implementation from the previous turn. If they still pass, the tests lack sufficient discriminative power. In this case, we generate verbal feedback to guide the Coder toward more distinctive tests. If they fail as expected, the distinctiveness criterion is satisfied, and we proceed to semantic alignment.

\paragraph{Semantic Alignment}
Finally, an \textit{Evaluator} 
checks whether the generated tests faithfully reflect the requirements of the current round. It assesses whether the expected test outcomes
are consistent with the given specification. For instance, it can detect \textbf{Naming/Signature Inconsistency}, where functions, classes, or return types in the tests do not match the requirements, and \textbf{Message Inconsistency}, where the tests assert natural language outputs not required by the requirements.
Based on this assessment, the \textit{Evaluator} issues one of the following decisions:
\begin{itemize}
    \item \textsc{RETAIN}: The test cases are valid and preserved.
    \item \textsc{TEST REFINE}: The test cases require refinement, with detailed feedback provided to the \textit{Coder}.
    \item \textsc{QUESTION REFINE}: The issue originates from ambiguity or error in the requirement itself and requires human intervention before regeneration.
\end{itemize}
In the cases of \textsc{TEST REFINE} or \textsc{QUESTION REFINE}, actionable feedback is supplied to support revision. When \textsc{QUESTION REFINE} is triggered, the process pauses for human reviewers to resolve the instruction ambiguity, after which generation resumes.

\begin{table}[]
\centering
\caption{Statistics of \tool. \#Char, and \#Line represent the average number of test cases, characters, and lines per question, respectively. For the repository level, the ``Question'' in \#Char and \#Line value includes the code context above the function.}
\label{tab:bench_statistics}
\resizebox{\columnwidth}{!}{%
\begin{tabular}{cccccccc|rrr|rrr}
\toprule
\multirow{2}{*}{\textbf{Level}} &
  \multirow{2}{*}{\textbf{Lanauge}} &
  \multirow{2}{*}{\textbf{\#Task}} &
  \multirow{2}{*}{\textbf{\#Repo}} &
  \multicolumn{4}{c}{\textbf{\#Question}} &
  \multicolumn{3}{c}{\textbf{\#Char}} &
  \multicolumn{3}{c}{\textbf{\#Line}} \\ \cmidrule{5-14} 
              &           &     &    & Min & Max & Avg. & Total & Question & Code    & Test    & Question & Code  & Test \\ \midrule
Function      & Python    & 142 & -  & 3   & 5   & 3.94 & 560  & 233.30    & 1,234.40  & 2,023.30  & 2.20   & 36.30  & 50.00   \\
Repository    & Python    & 89  & 41 & 3   & 5   & 4.24 & 377   & 11,495.10  & 605.70   & 1,074.50  & 333.70 & 18.20  & 28.70 \\
Function      & Java      & 172 & -  & 3  & 5   & 4.58 & 788   & 361.60    & 2,051.60 & 2,184.10 & 3.30     & 63.90 & 57.00  \\
Repository    & Java      & 40  & 5  & 2  & 5   & 3.30  & 132   & 6,753.70   & 1,169.40  & 1,140.80  & 148.30    & 28.90  & 25.90 \\ \midrule
\multicolumn{2}{c}{Total} & 443 & 46 & 2 & 5   & 4.19 & 1,857  & 270.80   & 1,419.50  & 1,815.40  & 2.30 & 42.70  & 46.20 \\ \bottomrule
\end{tabular}%
}
\end{table}


\subsection{Benchmark Statistics}\label{sec:benchmark_stat}

As shown in Table~\ref{tab:bench_statistics}, \tool covers both Python and Java, and supports two levels of evaluation granularity: function-level and repository-level. Each data sample in the benchmark represents a programming task spanning 2 to 5 turns. In each turn, the benchmark provides a triplet including a user requirement, a reference implementation, and test cases. It begins with an initial core requirement and evolves through requirements that add features, handle edge cases, or adjust the design. 

The benchmark comprises a total of 443 tasks. For Python, there are 231 questions split between 142 function-level tasks and 89 repository-level tasks spanning 41 repositories. For Java, there are 212 tasks divided into 172 function-level tasks and 40 repository-level tasks across 5 repositories. 
In total, these tasks generated 1,857 questions.
Each programming task involves an average of 4.19 interaction turns, with the number of turns ranging from a minimum of 2 to a maximum of 5, reflecting the real-world iterative development scenarios.

\section{Experiment Setup}\label{sec:experiment}

\subsection{Research Questions}\label{sec:experiment_rq}

Our experiments aim to answer the following research questions:
\begin{itemize}
    
    \item \textbf{RQ1: How do current LLMs perform in iterative code generation?}
    In this research question, we investigate the effectiveness of 11 LLMs across different sizes and types in iterative development scenarios.
    
    \item \textbf{RQ2: How do different prompting strategies affect the performance of LLMs in iterative code generation task?} In this research question, we evaluate three common prompting methods, including \textit{Full History}, \textit{Code Edit}, and \textit{Cumulative Instruction} to understand their impact on model performance.
    
    \item \textbf{RQ3: How effective of our test case generation strategy in enhancing test case's quality and discriminative capability?} 
    In this research question, we examine whether our Semantic-aware Discriminative Test Case Generation method enhances the quality and discriminative capability of test cases by comparing benchmark variants with and without different verification steps.
\end{itemize}

\definecolor{darkgreen}{rgb}{0,0.5,0}
\definecolor{lightblue}{RGB}{230,240,255}
\newcommand{\cross}{\textcolor{red}{\textbf{\XSolidBrush}}}
\newcommand{\tick}{\textcolor{darkgreen}{\Checkmark}}
\begin{table}[t] 
\centering
\caption{List of selected models, their categories, abbreviations, size, and the open-source status.}
\label{tab:evaluation_models}
\footnotesize
\scalebox{0.9}{
\begin{tabular}{lllcc}
\toprule
\rowcolor{gray!20} \textbf{Category} & \textbf{Model name} & \textbf{Abbreviation} & \textbf{Size} & \textbf{Open-source}\\
\midrule

\multirow{3}{*}{Large reasoning models} 
& \texttt{gpt-oss-120B}~\citep{openai2025gptoss120bgptoss20bmodel} & OSS-120-T & 120B & \tick \\
& \texttt{DeepSeek-V3.1-Think}~\citep{DBLP:journals/corr/abs-2501-12948} & DS-T & 370B & \tick \\
& \texttt{Qwen3-235B-A22B-Thinking-2507}~\citep{yang2025qwen3} & QW3-235-T & 235B & \tick \\
\midrule

\multirow{4}{*}{Large language models} 
& \texttt{GPT-5-mini}~\citep{gpt-5} & G5-M & - & \cross \\
& \texttt{Qwen3-Coder-480B-A35B-Instruct}~\citep{qwen3technicalreport} & QW3-Coder & 480B & \tick \\
& \texttt{Qwen3-235B-A22B-Instruct-2507}~\citep{yang2025qwen3} & QW3-235 & 235B & \tick \\
& \texttt{DeepSeek-V3.1}~\citep{liu2024deepseek} & DS & 370B & \tick \\
\midrule

\multirow{2}{*}{Small reasoning models} 
& \texttt{Qwen3-30B-A3B-Think}~\citep{yang2025qwen3} & QW3-30-T & 30B & \tick \\
& \texttt{gpt-oss-20B}~\citep{openai2025gptoss120bgptoss20bmodel} & OSS-20-T & 20B & \tick \\
\midrule

\multirow{2}{*}{Small language models} 
& \texttt{Qwen3-30B-A3B}~\citep{yang2025qwen3} & QW3-30 & 30B &  \tick \\
& \texttt{gpt-oss-20B}~\citep{openai2025gptoss120bgptoss20bmodel} & OSS-20 & 20B & \tick \\

\bottomrule
\end{tabular}
}
\end{table}

\subsection{Selected Models}\label{sec:experiment_model}
To provide a comprehensive and representative evaluation of LLMs in iterative development tasks, we select a diverse set of models from four distinct categories: Large Reasoning Models, Large Language Models, Small Reasoning Models, and Small Language Models.
Table~\ref{tab:evaluation_models} provides a detailed list of these models, including their abbreviations, parameter sizes, and open-source status.
Notably, the reasoning level for \texttt{gpt-oss-120B} and the reasoning version of \texttt{gpt-oss-20B} is set to ``Medium'', while the language model version of \texttt{gpt-oss-20B} is set to ``Low''.






\subsection{Prompt Design}\label{sec:experiment_prompt}

\begin{figure}[t]
    \centering
    \includegraphics[width=0.85\linewidth, keepaspectratio]{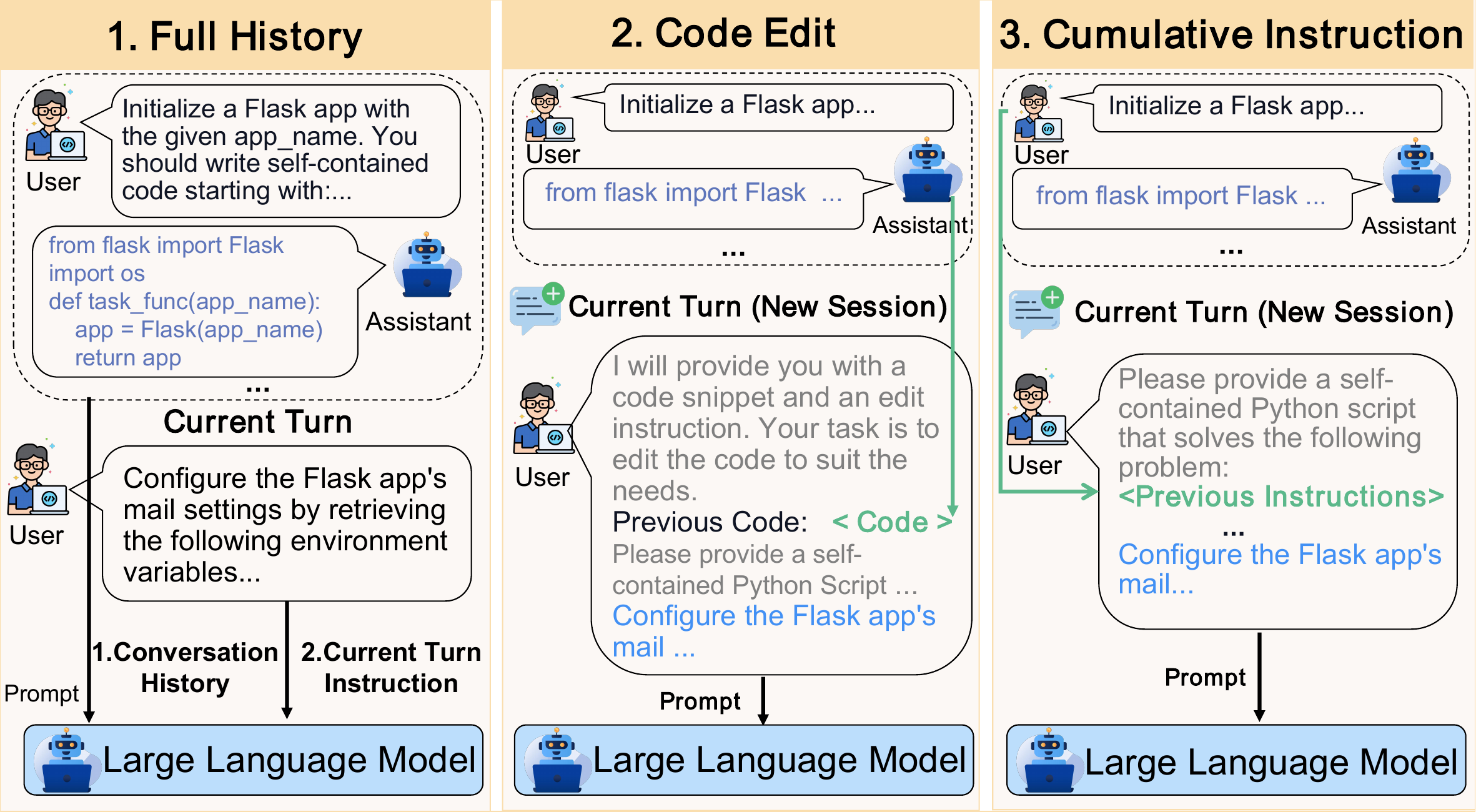}
    \caption{Three prompting strategies in iterative code generation.}
    \label{fig:prompt_design}
\end{figure}

\paragraph{Prompting Strategies}
To simulate realistic developer interaction patterns in iterative development scenarios, we design three {prompting} strategies as shown in Fig.~\ref{fig:prompt_design}, including \textbf{Full History}, \textbf{Code Edit}, and \textbf{Cumulative Instruction}. Each strategy represents a distinct mode of iterative interaction commonly observed when developers collaborate with AI-assisted programming tools:

\begin{itemize}
    \item \textbf{Full History}:
    In this setting, all preceding user instructions and model responses are included in the prompt. This simulates an uninterrupted multi-turn conversation thread, where the developer incrementally adds new requirements and expects the model to preserve and build upon the evolving context.

    \item \textbf{Code Edit}:
    The prompt {solely contains} 
    the code generated in the previous turn along with the new user instruction. This mimics a developer opening a fresh interaction window after receiving an initial solution, with the intent to refine, modify, or extend the existing codebase.

    \item \textbf{Cumulative Instruction}:
    This strategy accumulates all user instructions across turns but omits any intermediate model outputs. It reflects scenarios in which developers iteratively clarify and augment the task specification, without incorporating or reviewing prior code generations.
\end{itemize}

In repository-level evaluations, {we further follow the setting in DevEval and MRGBench to incorporate localized code context into the prompt.}

\paragraph{Prompt Context Settings}
In addition to prompt strategies, we also consider
two context settings to investigate the influence of historical code quality on model behavior during multi-turn interaction:

\begin{itemize}
    \item \textbf{Basic Setting}:
    In each turn, the prompt includes the model-generated code from the previous rounds (as dictated by the chosen prompt strategy). This reflects the natural accumulation of model outputs over time and simulates a realistic, possibly noisy, development process.

    \item \textbf{Golden Setting}:
    In contrast, the prompt is constructed using the ground-truth reference code from earlier rounds. It simulates the scenarios where the developer does not directly use the LLM's solution but manually checks and rectifies it. This setting serves as an upper-bound scenario, where the historical code context is assumed to be correct and complete.
\end{itemize}

\subsection{Evaluation Metrics}\label{sec:experiment_evaluation}

We employ execution-based metrics to quantify the performance of LLMs in iterative code generation. Our primary metric is \textbf{pass@1} \citep{DBLP:journals/corr/abs-2107-03374}, which measures whether the generated code passes the test suite on the first attempt. Based on it, we define the metrics used for iterative code generation as follows:
\begin{itemize}
    \item \textbf{Per-turn Accuracy} ($\text{Acc}^{(k)}$): the success rate of the model at each individual turn. 
    
    \item \textbf{Average Accuracy} (Avg Acc): the mean of the per-turn accuracies across all turns.
    
    \item \textbf{Complete Task Rate} (CR): the proportion of tasks where the model successfully completes all required turns. 

    \item \textbf{Average Token Cost} ($\text{ATC}$): the mean total number of tokens (input and output) consumed per task across all turns.
\end{itemize}

\subsection{Implementation Details}\label{sec:impl_detail}

\paragraph{Experimental Environment}

All experiments are conducted on a server equipped with dual Intel Xeon Gold 6240C CPUs (totaling 72 cores/144 threads), 156.8GB of memory, running a Linux distribution. 
For commercial models such as GPT-5.1 and DeepSeek variants, we use their official APIs. For gpt-oss and Qwen3 variants, we rely on the hosted versions available on the DeepInfra \citep{deepinfra} platform.

\paragraph{Data Synthesis}
 
To balance diversity and consistency in natural language outputs, we set the generation temperature of the LLMs behind the \textit{docstring converter}, \textit{decomposer}, \textit{analyzer}, and \textit{evaluator} (Section~\ref{sec:pipeline}) to 0.5, which allows for moderate variability while maintaining reliability. These components are powered by \textit{Qwen3-235B-A22B-Instruct-2507}. 
We set the temperature to 0 for the \textit{coder}, implemented with \textit{Claude-sonnet-4}, to maximize its coding capabilities.
The default maximum number of iterations in Section~\ref{sec:decomposition} is 6, and in Section~\ref{sec:codegen} it is 10. For the code execution environment, we adopt the same environment as the original seed dataset \citep{DBLP:conf/iclr/ZhuoVCH0WYZHPB025,
chou2025autocodebenchlargelanguagemodels,
DBLP:conf/acl/Li0ZLLZWLFWDZZD24} and have adapted it to support multi-turn scenarios.

\paragraph{Evaluation}

To eliminate the influence of random sampling, we utilize a greedy decoding strategy during inference.
For \textit{gpt-oss} variants, considering the costs and inference latency, we regard reasoning level ``medium'' as the thinking mode and reasoning level ``low'' as the non-thinking mode. All other model parameters are kept at their default values. 
\begin{table}[]
\centering
\caption{Performance of various LLMs on function-level tasks. 
    For each column, \textbf{bold} indicates the best performance across all models, and \underline{underlining} indicates the best performance within the same model group. 
    BS denotes the basic context setting, GS denotes the golden context setting, Avg Acc indicates the average accuracy for five rounds, and CR indicates the completion rate. 
    A full list of models and their abbreviations is provided in Table~\ref{tab:evaluation_models}. The same highlighting and abbreviations are also used in Tables~\ref{tab:rq2_repo}.}
\label{tab:rq2_func}
\resizebox{\columnwidth}{!}
{%
\begin{tabular}{lcccccccccccccc}
\toprule
{\color[HTML]{000000}} &
  \multicolumn{2}{c}{{\color[HTML]{000000} \textbf{Round 1}}} &
  \multicolumn{2}{c}{{\color[HTML]{000000} \textbf{Round 2}}} &
  \multicolumn{2}{c}{{\color[HTML]{000000} \textbf{Round 3}}} &
  \multicolumn{2}{c}{{\color[HTML]{000000} \textbf{Round 4}}} &
  \multicolumn{2}{c}{{\color[HTML]{000000} \textbf{Round 5}}} &
  \multicolumn{2}{c}{{\color[HTML]{000000} \textbf{Avg Acc}}} &
  \multicolumn{2}{c}{{\color[HTML]{000000} \textbf{CR}}} \\
  \cmidrule{2-15}
\multirow{-2}{*}{{\color[HTML]{000000} \textbf{Model}}} &
  {\color[HTML]{000000} BS} &
  {\color[HTML]{000000} GS} &
  {\color[HTML]{000000} BS} &
  {\color[HTML]{000000} GS} &
  {\color[HTML]{000000} BS} &
  {\color[HTML]{000000} GS} &
  {\color[HTML]{000000} BS} &
  {\color[HTML]{000000} GS} &
  {\color[HTML]{000000} BS} &
  {\color[HTML]{000000} GS} &
  {\color[HTML]{000000} BS} &
  {\color[HTML]{000000} GS} &
  {\color[HTML]{000000} BS} &
  {\color[HTML]{000000} GS} \\ \midrule
\rowcolor{gray!20} \multicolumn{15}{c}{\textbf{Python}} \\ \midrule
{\color[HTML]{000000} DS-T} &
  {\color[HTML]{000000} 0.4648} &
  {\color[HTML]{000000} 0.5211} &
  {\color[HTML]{000000} 0.5915} &
  {\color[HTML]{000000} 0.7324} &
  {\color[HTML]{000000} 0.4085} &
  {\color[HTML]{000000} 0.6549} &
  {\color[HTML]{000000} 0.4600} &
  {\color[HTML]{000000} 0.6300} &
  {\color[HTML]{000000} 0.4118} &
  {\color[HTML]{000000} 0.5882} &
  {\color[HTML]{000000} 0.4673} &
  {\color[HTML]{000000} 0.6253} &
  {\color[HTML]{000000} 0.1408} &
  {\color[HTML]{000000} 0.1831} \\
{\color[HTML]{000000} OSS-120-T} &
  {\color[HTML]{000000} {\ul 0.5282}} &
  {\color[HTML]{000000} {\ul 0.5563}} &
  {\color[HTML]{000000} {\ul 0.6056}} &
  {\color[HTML]{000000} {\ul \textbf{0.7535}}} &
  {\color[HTML]{000000} {\ul 0.4648}} &
  {\color[HTML]{000000} {\ul 0.6620}} &
  {\color[HTML]{000000} {\ul 0.4700}} &
  {\color[HTML]{000000} {\ul 0.6400}} &
  {\color[HTML]{000000} 0.4412} &
  {\color[HTML]{000000} {\ul 0.6765}} &
  {\color[HTML]{000000} 0.5020} &
  {\color[HTML]{000000} {\ul 0.6577}} &
  {\color[HTML]{000000} {\ul 0.1972}} &
  {\color[HTML]{000000} {\ul \textbf{0.2746}}} \\
{\color[HTML]{000000} QW3-235-T} &
  {\color[HTML]{000000} 0.4859} &
  {\color[HTML]{000000} 0.4648} &
  {\color[HTML]{000000} 0.5563} &
  {\color[HTML]{000000} {\ul \textbf{0.7535}}} &
  {\color[HTML]{000000} 0.4366} &
  {\color[HTML]{000000} 0.6549} &
  {\color[HTML]{000000} 0.4400} &
  {\color[HTML]{000000} 0.6200} &
  {\color[HTML]{000000} {\ul \textbf{0.6176}}} &
  {\color[HTML]{000000} 0.5882} &
  {\color[HTML]{000000} {\ul 0.5073}} &
  {\color[HTML]{000000} 0.6163} &
  {\color[HTML]{000000} 0.1549} &
  {\color[HTML]{000000} 0.1761} \\ \midrule
{\color[HTML]{000000} DS} &
  {\color[HTML]{000000} 0.5070} &
  {\color[HTML]{000000} 0.5282} &
  {\color[HTML]{000000} {\ul \textbf{0.6268}}} &
  {\color[HTML]{000000} {\ul \textbf{0.7535}}} &
  {\color[HTML]{000000} {\ul \textbf{0.4859}}} &
  {\color[HTML]{000000} 0.6408} &
  {\color[HTML]{000000} 0.4400} &
  {\color[HTML]{000000} {\ul \textbf{0.6700}}} &
  {\color[HTML]{000000} 0.4706} &
  {\color[HTML]{000000} 0.6765} &
  {\color[HTML]{000000} 0.5061} &
  {\color[HTML]{000000} 0.6538} &
  {\color[HTML]{000000} {\ul \textbf{0.2042}}} &
  {\color[HTML]{000000} 0.2042}\\
{\color[HTML]{000000} QW3-Coder} &
  {\color[HTML]{000000} {\ul \textbf{0.5704}}} &
  {\color[HTML]{000000} {\ul \textbf{0.5915}}} &
  {\color[HTML]{000000} 0.6197} &
  {\color[HTML]{000000} 0.7183} &
  {\color[HTML]{000000} 0.4718} &
  {\color[HTML]{000000} 0.6620} &
  {\color[HTML]{000000} {\ul \textbf{0.4800}}} &
  {\color[HTML]{000000} 0.6400} &
  {\color[HTML]{000000} 0.4706} &
  {\color[HTML]{000000} {\ul \textbf{0.7353}}} &
  {\color[HTML]{000000} 0.5225} &
  {\color[HTML]{000000} {\ul \textbf{0.6694}}} &
  {\color[HTML]{000000} 0.1972} &
  {\color[HTML]{000000} 0.2254} \\
{\color[HTML]{000000} G5-M} &
  {\color[HTML]{000000} 0.4577} &
  {\color[HTML]{000000} 0.4859} &
  {\color[HTML]{000000} 0.5493} &
  {\color[HTML]{000000} {\ul \textbf{0.7535}}} &
  {\color[HTML]{000000} 0.4225} &
  {\color[HTML]{000000} {\ul \textbf{0.7183}}} &
  {\color[HTML]{000000} 0.4400} &
  {\color[HTML]{000000} 0.5900} &
  {\color[HTML]{000000} 0.5000} &
  {\color[HTML]{000000} 0.6765} &
  {\color[HTML]{000000} 0.4739} &
  {\color[HTML]{000000} 0.6448} &
  {\color[HTML]{000000} 0.1901} &
  {\color[HTML]{000000} {\ul 0.2535}}\\
{\color[HTML]{000000} QW3-235} &
  {\color[HTML]{000000} 0.5423} &
  {\color[HTML]{000000} 0.5563} &
  {\color[HTML]{000000} 0.5986} &
  {\color[HTML]{000000} 0.7324} &
  {\color[HTML]{000000} 0.4718} &
  {\color[HTML]{000000} 0.6972} &
  {\color[HTML]{000000} 0.5200} &
  {\color[HTML]{000000} 0.6400} &
  {\color[HTML]{000000} {\ul 0.5294}} &
  {\color[HTML]{000000} 0.7059} &
  {\color[HTML]{000000} {\ul \textbf{0.5324}}} &
  {\color[HTML]{000000} 0.6664} &
  {\color[HTML]{000000} 0.1620} &
  {\color[HTML]{000000} 0.1972}\\ \midrule
{\color[HTML]{000000} QW3-30-T} &
  {\color[HTML]{000000} 0.4155} &
  {\color[HTML]{000000} 0.4577} &
  {\color[HTML]{000000} 0.4859} &
  {\color[HTML]{000000} 0.7042} &
  {\color[HTML]{000000} 0.4507} &
  {\color[HTML]{000000} {\ul 0.6761}} &
  {\color[HTML]{000000} {\ul 0.4700}} &
  {\color[HTML]{000000} 0.6400} &
  {\color[HTML]{000000} 0.3824} &
  {\color[HTML]{000000} 0.5882} &
  {\color[HTML]{000000} 0.4409} &
  {\color[HTML]{000000} 0.6133} &
  {\color[HTML]{000000} 0.1197} &
  {\color[HTML]{000000} 0.1549}\\
{\color[HTML]{000000} OSS-20-T} &
  {\color[HTML]{000000} {\ul 0.5352}} &
  {\color[HTML]{000000} {\ul 0.5423}} &
  {\color[HTML]{000000} {\ul 0.5986}} &
  {\color[HTML]{000000} {\ul 0.7324}} &
  {\color[HTML]{000000} {\ul 0.4718}} &
  {\color[HTML]{000000} 0.6479} &
  {\color[HTML]{000000} 0.4400} &
  {\color[HTML]{000000} {\ul 0.6600}} &
  {\color[HTML]{000000} {\ul 0.4118}} &
  {\color[HTML]{000000} {\ul 0.6176}} &
  {\color[HTML]{000000} {\ul 0.4915}} &
  {\color[HTML]{000000} {\ul 0.6400}} &
  {\color[HTML]{000000} {\ul 0.1620}} &
  {\color[HTML]{000000} {\ul 0.2183}} \\ \midrule
{\color[HTML]{000000} QW3-30} &
  {\color[HTML]{000000} 0.4155} &
  {\color[HTML]{000000} 0.4155} &
  {\color[HTML]{000000} 0.4507} &
  {\color[HTML]{000000} 0.6761} &
  {\color[HTML]{000000} 0.3662} &
  {\color[HTML]{000000} 0.5634} &
  {\color[HTML]{000000} 0.3200} &
  {\color[HTML]{000000} 0.5700} &
  {\color[HTML]{000000} 0.3824} &
  {\color[HTML]{000000} {\ul 0.6765}} &
  {\color[HTML]{000000} 0.3869} &
  {\color[HTML]{000000} 0.5803} &
  {\color[HTML]{000000} 0.0845} &
  {\color[HTML]{000000} 0.1056}\\
{\color[HTML]{000000} OSS-20} &
  {\color[HTML]{000000} {\ul 0.5352}} &
  {\color[HTML]{000000} {\ul 0.5423}} &
  {\color[HTML]{000000} {\ul 0.5493}} &
  {\color[HTML]{000000} {\ul 0.7183}} &
  {\color[HTML]{000000} {\ul 0.4648}} &
  {\color[HTML]{000000} {\ul 0.6338}} &
  {\color[HTML]{000000} {\ul 0.3900}} &
  {\color[HTML]{000000} {\ul 0.6500}} &
  {\color[HTML]{000000} {\ul 0.5294}} &
  {\color[HTML]{000000} {\ul 0.6765}} &
  {\color[HTML]{000000} {\ul 0.4937}} &
  {\color[HTML]{000000} {\ul 0.6442}} &
  {\color[HTML]{000000} {\ul 0.1901}} &
  {\color[HTML]{000000} {\ul 0.1972}} \\

  \midrule
\rowcolor{gray!20} \multicolumn{15}{c}{\textbf{Java}} \\ \midrule
DS-T &
  {\ul 0.6163} &
  {\ul 0.6105} &
  {\ul 0.5814} &
  {\ul 0.6628} &
  0.4767 &
  {\ul 0.7035} &
  {\ul 0.4847} &
  0.6258 &
  {\ul 0.4679} &
  {\ul 0.6055} &
  {\ul 0.5254} &
  {\ul 0.6416} &
  {\ul 0.1977} &
  {\ul 0.2267} \\
OSS-120-T &
  0.5756 &
  {\ul 0.6105} &
  0.5291 &
  0.6337 &
  0.4651 &
  0.6453 &
  {\ul 0.4847} &
  {\ul 0.6380} &
  0.4312 &
  0.5688 &
  0.4971 &
  0.6193 &
  {\ul 0.1977} &
  {\ul 0.2267} \\
QW3-235-T &
  0.5814 &
  0.5465 &
  0.5349 &
  0.6570 &
  {\ul 0.5000} &
  0.6047 &
  0.4540 &
  0.4724 &
  {\ul 0.4679} &
  0.3853 &
  0.5076 &
  0.5332 &
  0.1744 &
  0.1337 \\ \midrule
DS &
  0.6221 &
  0.6105 &
  0.5698 &
  0.6919 &
  0.5233 &
  {\ul \textbf{0.7267}} &
  {\ul 0.4908} &
  0.6933 &
  0.4679 &
  0.6239 &
  0.5348 &
  0.6692 &
  {\ul \textbf{0.2267}} &
  {\ul \textbf{0.2384}} \\
QW3-Coder &
  {\ul \textbf{0.7093}} &
  {\ul \textbf{0.7035}} &
  {\ul \textbf{0.5930}} &
  0.6919 &
  {\ul \textbf{0.5523}} &
  0.7035 &
  {\ul 0.4908} &
  0.6564 &
  {\ul 0.4771} &
  {\ul \textbf{0.6606}} &
  {\ul \textbf{0.5645}} &
  {\ul \textbf{0.6832}} &
  0.1919 &
  0.2209 \\
G5-M &
  0.5349 &
  0.5465 &
  0.5058 &
  0.6802 &
  0.4535 &
  0.6802 &
  {\ul 0.4908} &
  {\ul \textbf{0.7117}} &
  0.4404 &
  0.6514 &
  0.4851 &
  0.6540 &
  0.2035 &
  0.2267 \\
QW3-235 &
  0.6512 &
  0.6512 &
  0.5698 &
  {\ul \textbf{0.6977}} &
  0.4709 &
  0.6919 &
  0.4110 &
  0.6626 &
  {\ul 0.4771} &
  0.6055 &
  0.5160 &
  0.6618 &
  0.1570 &
  0.2209 \\ \midrule
{\color[HTML]{000000} QW3-30-T} &
  0.5291 &
  0.5465 &
  0.4419 &
  0.6047 &
  0.4012 &
  0.5872 &
  0.3865 &
  {\ul 0.6564} &
  0.3211 &
  {\ul 0.5505} &
  0.4159 &
  0.5891 &
  0.1047 &
  0.1279 \\
OSS-20-T &
  {\ul 0.5581} &
  {\ul 0.5640} &
  {\ul 0.5640} &
  {\ul 0.6453} &
  {\ul 0.5174} &
  {\ul 0.6860} &
  {\ul \textbf{0.4969}} &
  0.6442 &
  {\ul \textbf{0.4954}} &
  0.5229 &
  {\ul 0.5264} &
  {\ul 0.6125} &
  {\ul 0.1977} &
  {\ul 0.1977} \\ \midrule
{\color[HTML]{000000} QW3-30} &
  0.5291 &
  0.5233 &
  0.4709 &
  0.5872 &
  0.3953 &
  0.6047 &
  0.3742 &
  0.5460 &
  0.2569 &
  0.5138 &
  0.4053 &
  0.5550 &
  0.0581 &
  0.0872 \\
OSS-20 &
  {\ul 0.5988} &
  {\ul 0.6105} &
  {\ul 0.5756} &
  {\ul 0.6802} &
  {\ul 0.5116} &
  {\ul 0.6744} &
  {\ul 0.4908} &
  {\ul 0.6258} &
  {\ul 0.4771} &
  {\ul 0.5872} &
  {\ul 0.5308} &
  {\ul 0.6356} &
  {\ul 0.2035} &
  {\ul 0.2151} \\ \bottomrule
\end{tabular}%
}
\end{table}

\begin{table}[]
\centering
\caption{Performance of various LLMs on repository-level task.}
\label{tab:rq2_repo}
\resizebox{\columnwidth}{!}{%
\begin{tabular}{lcccccccccccccc}
\toprule
 &
  \multicolumn{2}{c}{\textbf{Round 1}} &
  \multicolumn{2}{c}{\textbf{Round 2}} &
  \multicolumn{2}{c}{\textbf{Round 3}} &
  \multicolumn{2}{c}{\textbf{Round 4}} &
  \multicolumn{2}{c}{\textbf{Round 5}} &
  \multicolumn{2}{c}{\textbf{Avg Acc}} &
  \multicolumn{2}{c}{\textbf{CR}}\\
  \cmidrule{2-15}
\multirow{-2}{*}{\textbf{Model}} &
  BS &
  GS &
  BS &
  GS &
  BS &
  GS &
  BS &
  GS &
  BS &
  GS &
  BS &
  GS &
  BS &
  GS \\ \midrule
\rowcolor{gray!20} \multicolumn{15}{c}{\textbf{Python}} \\ \midrule
DS-T &
  0.2360 &
  0.2360 &
  {\ul 0.3371} &
  {\ul 0.4494} &
  {\ul 0.3146} &
  {\ul 0.4045} &
  {\ul 0.4588} &
  {\ul 0.6000} &
  0.3200 &
  {\ul \textbf{0.4800}} &
  {\ul 0.3333} &
  {\ul 0.4340} &
  0.0225 &
  {\ul 0.0674} \\
OSS-120-T &
  {\ul 0.2697} &
  {\ul 0.2584} &
  0.2921 &
  0.3371 &
  0.2809 &
  0.3483 &
  0.4235 &
  0.5294 &
  {\ul 0.4000} &
  0.4000 &
  0.3332 &
  0.3746 &
  {\ul 0.0449} &
  0.0337 \\
QW3-235-T &
  0.1011 &
  0.1011 &
  0.2135 &
  0.2472 &
  0.2472 &
  0.2584 &
  0.3412 &
  0.3294 &
  0.2800 &
  0.2800 &
  0.2366 &
  0.2432 &
  0.0225 &
  0.0225 \\ \midrule
DS &
  0.2360 &
  0.2360 &
  0.3596 &
  0.4157 &
  0.3146 &
  {\ul \textbf{0.4157}} &
  0.3882 &
  0.5647 &
  0.4000 &
  0.4000 &
  0.3397 &
  0.4064 &
  0.0562 &
  0.0562 \\
QW3-Coder &
  0.2135 &
  0.2584 &
  0.3483 &
  {\ul \textbf{0.5281}} &
  0.2921 &
  0.3820 &
  0.4353 &
  {\ul \textbf{0.6353}} &
  0.2000 &
  {\ul \textbf{0.4800}} &
  0.2978 &
  {\ul \textbf{0.4568}} &
  0.0225 &
  0.0449 \\
G5-M &
  {\ul \textbf{0.3483}} &
  {\ul \textbf{0.3034}} &
  {\ul \textbf{0.3933}} &
  0.4157 &
  {\ul \textbf{0.3483}} &
  0.3820 &
  {\ul \textbf{0.4824}} &
  0.6235 &
  0.4000 &
  0.4000 &
  {\ul \textbf{0.3945}} &
  0.4240 &
  {\ul \textbf{0.0674}} &
  {\ul \textbf{0.0787}} \\
QW3-235 &
  0.2474 &
  0.2584 &
  0.2921 &
  0.4494 &
  0.2921 &
  0.3708 &
  0.3882 &
  0.5882 &
  {\ul \textbf{0.4400}} &
  {\ul \textbf{0.4800}} &
  0.3319 &
  0.4294 &
  0.0449 &
  0.0449 \\ \midrule
{\color[HTML]{000000} QW3-30-T} &
  0.2022 &
  0.1910 &
  {\ul 0.3146} &
  {\ul 0.3596} &
  0.2247 &
  0.2809 &
  0.3059 &
  {\ul 0.5294} &
  0.2400 &
  0.3200 &
  0.2575 &
  0.3362 &
  0.0112 &
  0.0337 \\
OSS-20-T &
  {\ul 0.2614} &
  {\ul 0.2500} &
  0.2727 &
  0.3523 &
  {\ul 0.2614} &
  {\ul 0.3295} &
  {\ul 0.381} &
  0.5238 &
  {\ul 0.3200} &
  {\ul 0.4583} &
  {\ul 0.2993} &
  {\ul 0.3828} &
  {\ul 0.0337} &
  {\ul 0.0449} \\ \midrule
{\color[HTML]{000000} QW3-30} &
  0.2022 &
  0.1910 &
  {\ul 0.2809} &
  0.3371 &
  0.2022 &
  0.3146 &
  0.3294 &
  0.5529 &
  {\ul 0.3200} &
  {\ul \textbf{0.4800}} &
  0.2669 &
  0.3751 &
  {\ul 0.0225} &
  {\ul 0.0449} \\
OSS-20 &
  {\ul 0.2472} &
  {\ul 0.2584} &
  {\ul 0.2809} &
  {\ul 0.4045} &
  {\ul 0.2584} &
  {\ul 0.3371} &
  {\ul 0.3765} &
  {\ul 0.5647} &
  0.2000 &
  0.3600 &
  {\ul 0.2726} &
  {\ul 0.3849} &
  0.0112 &
  {\ul 0.0449} \\ \midrule
\rowcolor{gray!20} \multicolumn{15}{c}{\textbf{Java}} \\ \midrule
DS-T &
  0.3750 &
  {\ul 0.4250} &
  0.2000 &
  0.3750 &
  0.1515 &
  0.4545 &
  0.0667 &
  {\ul 0.4667} &
  0.0000 &
  {\ul \textbf{0.7500}} &
  0.1586 &
  {\ul 0.4942} &
  0.1000 &
  0.1000 \\
OSS-120-T &
  0.2500 &
  0.2250 &
  0.0500 &
  0.0250 &
  0.0606 &
  0.0606 &
  0.0000 &
  0.0000 &
  0.0000 &
  0.0000 &
  0.0721 &
  0.0621 &
  0.0250 &
  0.0000 \\
QW3-235-T &
  {\ul 0.4000} &
  0.4000 &
  {\ul \textbf{0.3000}} &
  {\ul 0.4250} &
  {\ul 0.2727} &
  {\ul 0.5152} &
  {\ul \textbf{0.1333}} &
  {\ul 0.4667} &
  0.0000 &
  0.0000 &
  {\ul 0.2212} &
  0.3614 &
  {\ul 0.2000} &
  {\ul 0.1500} \\ \midrule
DS &
  0.4000 &
  0.4750 &
  0.2750 &
  0.3500 &
  0.2727 &
  0.3636 &
  0.0667 &
  0.6000 &
  0.0000 &
  {\ul \textbf{0.7500}} &
  0.2029 &
  0.5077 &
  0.1500 &
  0.1000 \\
QW3-Coder &
  0.4000 &
  0.4000 &
  {\ul \textbf{0.3000}} &
  0.4000 &
  {\ul \textbf{0.3030}} &
  {\ul \textbf{0.6970}} &
  {\ul \textbf{0.1333}} &
  {\ul \textbf{0.9333}} &
  0.0000 &
  {\ul \textbf{0.7500}} &
  0.2273 &
  0.6361 &
  0.1500 &
  0.2000 \\
G5-M &
  0.2750 &
  0.3000 &
  0.1750 &
  0.2250 &
  0.1515 &
  0.3333 &
  {\ul \textbf{0.1333}} &
  0.2667 &
  0.0000 &
  0.2500 &
  0.1470 &
  0.2750 &
  0.1250 &
  0.0750 \\
QW3-235 &
  {\ul \textbf{0.5000}} &
  {\ul \textbf{0.5000}} &
  {\ul \textbf{0.3000}} &
  {\ul 0.4750} &
  {\ul \textbf{0.3030}} &
  0.6667 &
  0.0667 &
  0.8667 &
  0.0000 &
  {\ul \textbf{0.7500}} &
  {\ul \textbf{0.2339}} &
  {\ul \textbf{0.6517}} &
  {\ul \textbf{0.2000}} &
  {\ul \textbf{0.2250}} \\ \midrule
{\color[HTML]{000000} QW3-30-T} &
  0.3000 &
  0.2500 &
  {\ul 0.2250} &
  {\ul 0.4000} &
  {\ul 0.2121} &
  {\ul 0.5455} &
  {\ul \textbf{0.1333}} &
  {\ul 0.8000} &
  0.0000 &
  {\ul \textbf{0.7500}} &
  {\ul 0.1741} &
  {\ul 0.5491} &
  {\ul 0.1250} &
  {\ul 0.1000} \\
OSS-20-T &
  {\ul 0.3250} &
  {\ul 0.2750} &
  0.2000 &
  0.1750 &
  0.1515 &
  0.3030 &
  0.0667 &
  0.4000 &
  0.0000 &
  {\ul \textbf{0.7500}} &
  0.1486 &
  0.3806 &
  0.1000 &
  {\ul 0.1000} \\ \midrule
{\color[HTML]{000000} QW3-30} &
  0.2500 &
  0.2250 &
  {\ul \textbf{0.3000}} &
  {\ul \textbf{0.5000}} &
  {\ul \textbf{0.3030}} &
  {\ul 0.6061} &
  {\ul \textbf{0.1333}} &
  {\ul 0.8000} &
  0.0000 &
  {\ul \textbf{0.7500}} &
  {\ul 0.1973} &
  {\ul 0.5762} &
  {\ul 0.1250} &
  {\ul 0.1500} \\
OSS-20 &
  {\ul 0.3250} &
  {\ul 0.2750} &
  0.2250 &
  0.1750 &
  0.1818 &
  0.4242 &
  0.0667 &
  0.3333 &
  0.0000 &
  0.5000 &
  0.1597 &
  0.3415 &
  {\ul 0.1250} &
  0.1250 \\ \bottomrule
\end{tabular}%
}
\end{table}

\section{Results}\label{sec:result}

\subsection{RQ1: Overall Performance}\label{sec:result_rq1}

Table~\ref{tab:rq2_func} and Table~\ref{tab:rq2_repo} present the overall performance of eleven LLMs on the function-level and repository-level tasks of \tool, respectively.
From these results, we achieve the following observations.

\textbf{LLMs generally exhibit suboptimal performance in iterative code generation tasks, indicating that such scenarios remain a great challenge.}
In the basic context setting, which simulates realistic iterative interactions, the leading model's performance was modest. For function-level tasks, it achieved an average accuracy of just 53.24\% for Python and 56.45\% for Java, with task completion rates of only 20.42\% and 22.67\%, respectively. Performance on repository-level tasks was similarly limited, reaching an accuracy of 39.45\% for Python and 23.39\% for Java, with task completion rates of 6.74\% and 20.00\%.

Notably, under the golden context setting, where previous turns are perfectly resolved, performance improved greatly compared to the basic setting. For function-level tasks, accuracy improved to 66.94\% for Python and 68.32\% for Java, with corresponding completion rates of 27.46\% and 23.84\%. For repository-level tasks, the model achieved an accuracy of 45.68\% for Python and 65.17\% for Java, and a task completion rate of 7.87\% and 22.50\%, respectively.

These results indicate that iterative code generation remains a substantial challenge for current LLMs. The  performance gap between the basic and golden setting suggests that this challenge may stems from the need to maintain cumulative context and correct errors across iterations, which further underscores the importance of integrating code review into this development process to mitigate error accumulation.

\begin{tcolorbox}
\textbf{Finding 1:} LLMs demonstrate considerable limitations in iterative code generation. Systematic code review is essential to mitigate error accumulation during collaborative development with LLMs.
\end{tcolorbox}


\textbf{Model scale consistently improves performance in iterative code generation tasks, while reasoning mechanisms do not always provide an advantage.}
As shown in Table~\ref{tab:rq2_func} and Table~\ref{tab:rq2_repo},
a larger model Qwen3-235B-A22B-Instruct-2507 outperforms its smaller counterpart Qwen3-30B-A3B on function-level Python tasks with an average accuracy of 53.24\% vs. 38.69\%, and achieves similar gains on Java tasks with 51.60\% vs. 40.53\%. 
Comparable trends are observed at the repository level, with the larger model consistently surpassing the smaller one with Python accuracy 33.19\% vs 26.69\% and Java accuracy 23.39\% vs 19.73\%.

Additionally, models incorporating reasoning mechanisms do not consistently outperform their non-reasoning counterparts.
For instance, in function-level tasks, Deepseek-V3.1-think achieves an accuracy of 46.73\%, compared to 50.61\% for the non-reasoning version in Python tasks  and 52.54\% versus 53.48\% for Java tasks. In repository-level tasks, the performance gap persists: the reasoning version attains 33.33\% versus 33.97\% for non-reasoning version in Python, and 15.86\% versus 20.29\% in Java.
These observations suggest that reasoning mechanisms do not always provide an advantage in iterative code generation, contrasting with findings from competitive programming benchmarks~\citep{DBLP:conf/iclr/JainHGLYZWSSS25}. This discrepancy may be attributed to the phenomenon of overthinking mentioned in existing studies~\citep{DBLP:journals/tmlr/SuiCWZZYLWZZCH25,DBLP:journals/corr/abs-2504-13367}.

\begin{tcolorbox}
\textbf{Finding 2:} Incorporating reasoning capabilities does not consistently enhance model performance in iterative code generation tasks.
\end{tcolorbox}




\begin{table}[]
\setlength{\tabcolsep}{3.7mm}
\centering
\caption{Performance of different types of LLMs under different prompt strategies on function-level tasks. For each column, \textbf{bold} indicates the best
performance within the same context setting group. FH denotes Full History, CE denotes Code Edit, and CI denotes Cumulative Instruction. The same highlighting and abbreviations are also used in Tables~\ref{tab:rq3_repo}.}
\label{tab:rq3_func}
\footnotesize

\scalebox{0.85}{
\begin{tabular}{ccccccccc}
\toprule
\multirow{2}{*}{\textbf{Model}} &
  \multicolumn{1}{l}{\multirow{2}{*}{\textbf{Context}}} &
  \multicolumn{1}{l}{\multirow{2}{*}{\textbf{Prompting}}} &
  \multicolumn{3}{c}{\textbf{Python}} &
  \multicolumn{3}{c}{\textbf{Java}} \\ \cmidrule{4-9} 
 &
  \multicolumn{1}{l}{} &
  \multicolumn{1}{l}{} &
  \textbf{Avg Acc} &
  \textbf{CR} &
  \textbf{ATC$\downarrow$} &
  \textbf{Avg Acc} &
  \textbf{CR} &
  \textbf{ATC$\downarrow$} \\ \midrule
\multirow{5}{*}{DS-T}      & \multirow{3}{*}{BS} & FH & 0.4673          & 0.1408          & 7174.21          & 0.5254          & 0.1977          & 10,682.97         \\
                           &                     & CE & \textbf{0.4779} & \textbf{0.1479} & \textbf{6754.01} & 0.5027          & \textbf{0.2035} & 9664.94          \\
                           &                     & CI & 0.4659          & 0.1268          & 6889.74          & \textbf{0.5281} & 0.1570           & \textbf{9066.63} \\ \cmidrule{2-9} 
                           & \multirow{2}{*}{GS} & FH & 0.6253          & \textbf{0.1831} & \textbf{6943.65} & 0.6416          & 0.2267          & 10,259.47         \\
                           &                     & CE & \textbf{0.6303} & 0.1549          & 7012.32          & \textbf{0.6504} & \textbf{0.2326} & \textbf{9510.96} \\ \midrule
\multirow{5}{*}{QW3-Coder} & \multirow{3}{*}{BS} & FH & \textbf{0.5225} & 0.1972          & 3974.67          & 0.5645          & 0.1919          & 6998.23          \\
                           &                     & CE & 0.5124          & \textbf{0.2113} & 2790.33          & \textbf{0.5817} & \textbf{0.2267} & 4773.41          \\
                           &                     & CI & 0.5104          & 0.1338          & \textbf{2040.10}  & 0.5802          & 0.2093          & \textbf{3620.10}  \\ \cmidrule{2-9} 
                           & \multirow{2}{*}{GS} & FH & \textbf{0.6694} & \textbf{0.2254} & 3589.92          & 0.6832          & 0.2209          & 6115.76          \\
                           &                     & CE & 0.6526          & 0.2042          & \textbf{2567.62} & \textbf{0.6934} & \textbf{0.2384} & \textbf{4133.23} \\ \midrule
\multirow{5}{*}{OSS-20-T} &
  \multirow{3}{*}{BS} &
  FH &
  0.4915 &
  0.1620 &
  8358.19 &
  \textbf{0.5264} &
  \textbf{0.1977} &
  11,904.04 \\
                           &                     & CE & 0.4915          & \textbf{0.2183} & 7274.36          & 0.5068          & 0.1512          & 9881.18          \\
                           &                     & CI & \textbf{0.5210}  & 0.1761          & \textbf{6434.58} & 0.5242          & 0.1802          & \textbf{9098.91} \\ \cmidrule{2-9} 
                           & \multirow{2}{*}{GS} & FH & 0.6400            & \textbf{0.2183} & 7111.46          & 0.6125          & 0.1977          & 10,297.75         \\
                           &                     & CE & \textbf{0.6500}   & 0.2113          & \textbf{6310.47} & \textbf{0.6641} & \textbf{0.2093} & \textbf{8535.00}  \\ \midrule
\multirow{5}{*}{OSS-20}    & \multirow{3}{*}{BS} & FH & 0.4937          & 0.1901          & 8041.03          & \textbf{0.5308} & 0.2035          & 11,542.47         \\
                           &                     & CE & 0.5060           & \textbf{0.2113} & 6997.44          & 0.5212          & \textbf{0.2093} & 9664.45          \\
                           &                     & CI & \textbf{0.5321} & 0.1972          & \textbf{6581.85} & 0.5164          & 0.1744          & \textbf{8484.03} \\ \cmidrule{2-9} 
                           & \multirow{2}{*}{GS} & FH & 0.6442          & 0.1972          & 7201.30           & 0.6356          & \textbf{0.2151} & 10,271.83         \\
                           &                     & CE & \textbf{0.6627} & \textbf{0.2324} & \textbf{5631.84} & \textbf{0.6565} & \textbf{0.2151} & \textbf{7700.47} \\ \bottomrule
\end{tabular}%
}
\end{table}

\begin{table}[]
\setlength{\tabcolsep}{3.7mm}
\centering
\caption{Performance of different types of LLMs under different prompt strategies on repository-level tasks.}
\label{tab:rq3_repo}
\footnotesize
\scalebox{0.85}{
\begin{tabular}{ccccccccc}
\toprule
\multirow{2}{*}{\textbf{Model}} & \multirow{2}{*}{\textbf{Context}} & \multirow{2}{*}{\textbf{Prompting}} & \multicolumn{3}{c}{\textbf{Python}}          & \multicolumn{3}{c}{\textbf{Java}} \\ \cmidrule{4-9} 
                           &                     &    & \textbf{Avg Acc}         & \textbf{CR}              & \textbf{ATC$\downarrow$}   & \textbf{Avg Acc}         & \textbf{CR}             & \textbf{ATC$\downarrow$}   \\ \midrule
\multirow{5}{*}{DS-T}      & \multirow{3}{*}{BS} & FH & 0.3333          & 0.0225          & 25,856.01          & 0.1586          & 0.1000            & 10,940.27          \\
                           &                     & CE & 0.3240           & \textbf{0.0674} & \textbf{24,837.29} & \textbf{0.2279} & \textbf{0.1500}  & 11,183.55          \\
                           &                     & CI & \textbf{0.3696} & 0.0449          & 26,844.37          & 0.1436          & 0.0250          & \textbf{10,670.42} \\ \cmidrule{2-9} 
                           & \multirow{2}{*}{GS} & FH & \textbf{0.4340}  & \textbf{0.0674} & \textbf{22,982.36} & 0.4942          & 0.1000            & 10,504.62          \\
                           &                     & CE & 0.3922          & 0.0562          & 23,968.87          & \textbf{0.6492} & \textbf{0.2250} & \textbf{9731.38}  \\ \midrule
\multirow{5}{*}{QW3-Coder} & \multirow{3}{*}{BS} & FH & 0.2978          & 0.0225          & 15,444.08          & 0.2273          & 0.1500           & 4687.27           \\
                           &                     & CE & 0.3215          & 0.0337          & 14,311.97          & \textbf{0.2877} & \textbf{0.3250} & 3870.53           \\
                           &                     & CI & \textbf{0.3301} & \textbf{0.0449} & \textbf{13,304.37} & 0.1171          & 0.0000              & \textbf{3407.62}
                           \\ \cmidrule{2-9} 
                           & \multirow{2}{*}{GS} & FH & \textbf{0.4568} & \textbf{0.0449} & 13,939.88          & 0.6361          & 0.2000            & 4217.05           \\
                           &                     & CE & 0.3852          & 0.0337          & \textbf{13,344.83} & \textbf{0.7021} & \textbf{0.2750} & \textbf{3429.05}  \\ \midrule
\multirow{5}{*}{OSS-20-T}       & \multirow{3}{*}{BS}      & FH                         & 0.2993 & \textbf{0.0337} & \textbf{26,532.24} & 0.1486  & \textbf{0.1000}  & 12,452.20 \\
                           &                     & CE & 0.1945          & 0.0000               & 27,759.62          & \textbf{0.1547} & \textbf{0.1000}   & \textbf{9923.02}  \\
                           &                     & CI & \textbf{0.2998} & \textbf{0.0337} & 32,737.27          & 0.0871          & 0.0250          & 9955.62           \\ \cmidrule{2-9} 
                           & \multirow{2}{*}{GS} & FH & \textbf{0.3828} & \textbf{0.0449} & 26,893.75          & 0.3806          & 0.1000            & \textbf{9935.62}  \\
                           &                     & CE & 0.3296          & 0.0225          & \textbf{24,894.89} & \textbf{0.5329} & \textbf{0.1250} & 10,030.35          \\ \midrule
\multirow{5}{*}{OSS-20}    & \multirow{3}{*}{BS} & FH & 0.2726          & 0.0112          & 35,455.91          & 0.1597          & 0.1250          & 10,349.10           \\
                           &                     & CE & 0.1825          & 0.0000               & \textbf{32,018.91} & \textbf{0.1879} & \textbf{0.1750} & 9878.95           \\
                           &                     & CI & \textbf{0.3047} & \textbf{0.0449} & 46,583.43          & 0.1092          & 0.0250          & \textbf{9520.65}  \\ \cmidrule{2-9} 
                           & \multirow{2}{*}{GS} & FH & \textbf{0.3849} & \textbf{0.0449} & 33,246.78          & 0.3415          & 0.1250          & 14,824.10           \\
                           &                     & CE & 0.3034          & 0.0225         & \textbf{28,768.64} & \textbf{0.6011} & \textbf{0.2000}   & \textbf{11,701.83} \\ \bottomrule
\end{tabular}%
}
\end{table}


\subsection{RQ2: Prompt Strategy}\label{sec:result_rq2}

In this research question, we investigate the impact of different prompt strategies (as defined in Section~\ref{sec:experiment_prompt}) within iterative development scenarios. For this analysis, we selected the top-performing model on average from each category. For the golden setting, our investigation focused exclusively on the \textit{Full History} and \textit{Code Edit} strategies. The results are presented in Table~\ref{tab:rq3_func} and Table~\ref{tab:rq3_repo}.

\textbf{Prompt strategy is a critical determinant of both performance and cost-effectiveness in iterative code generation.} Our results reveal that \textit{Code Edit} (CE) generally achieves competitive performance and offers a better accuracy-token trade-off than \textit{Full History} (FH). Notably, at the repository level, \textit{Cumulative Instruction} (CI) occasionally outperforms CE in accuracy.
For instance, in function-level tasks using Qwen3-Coder-480B-A35B-Instruct, CE reduces token costs by 1,184.34 with a 1.41\% improvement in completion rate for Python, and by 2,224.82 tokens with 3.48\% gain for Java.
In repository-level tasks, CE achieves a complete rate improvements of 1.12\% with token savings of 1,132.11\% for Python, and a gain of 17.5\% with a reduction of 816.74\% tokens for Java.

These results can be explained by the fact that \textit{Full History} (FH) includes untrimmed context from all prior turns, which leads to excessive token consumption without proportionate accuracy gains. In contrast, CE and CI act as lightweight yet focused strategies that retain task-relevant information while discarding redundant history. This highlights the importance of \textbf{context engineering}~\citep{DBLP:journals/corr/abs-2507-13334} in iterative code generation, which plays a crucial role in balancing informativeness and efficiency in prolonged interactions with LLMs.

\begin{tcolorbox}

\textbf{Finding 3:} \textit{Code Edit} consistently offers a better balance of accuracy and token efficiency than \textit{Full History} in iterative code generation, while \textit{Cumulative Instruction} occasionally achieves higher accuracy at the repository level tasks. This highlights the critical role of context engineering in optimizing both performance and cost.

\end{tcolorbox}

\subsection{RQ3: Effectiveness of Semantic-Aware Discriminative Test Case Generation}\label{sec:result_rq3}
In this research question, we investigate whether our \textit{semantic-aware discriminative test case generation} method enhances test case quality, specifically by producing more robust and targeted test suites.

To this end, we conducted function-level experiments to minimize confounding variables from complex code contexts, thereby isolating the core impact of our validation strategy. Our experimental setup compared three distinct settings:
\begin{itemize}
    \item \textit{Ours}: The full approach, incorporating both the distinctiveness validation  and the semantic alignment module.
    \item \textit{w/o Evaluator}: Ablates the semantic alignment while retaining distinctiveness validation.
    \item \textit{w/o Distinctiveness Validation \& Evaluator}: Disables both modules.
\end{itemize}

We evaluated the performance of four representative LLMs on the datasets generated by each setting. In this context, a lower model accuracy indicates higher discriminative power and robustness of the generated test cases.

\begin{figure}[t]
    \centering
    \includegraphics[width=0.75\linewidth]{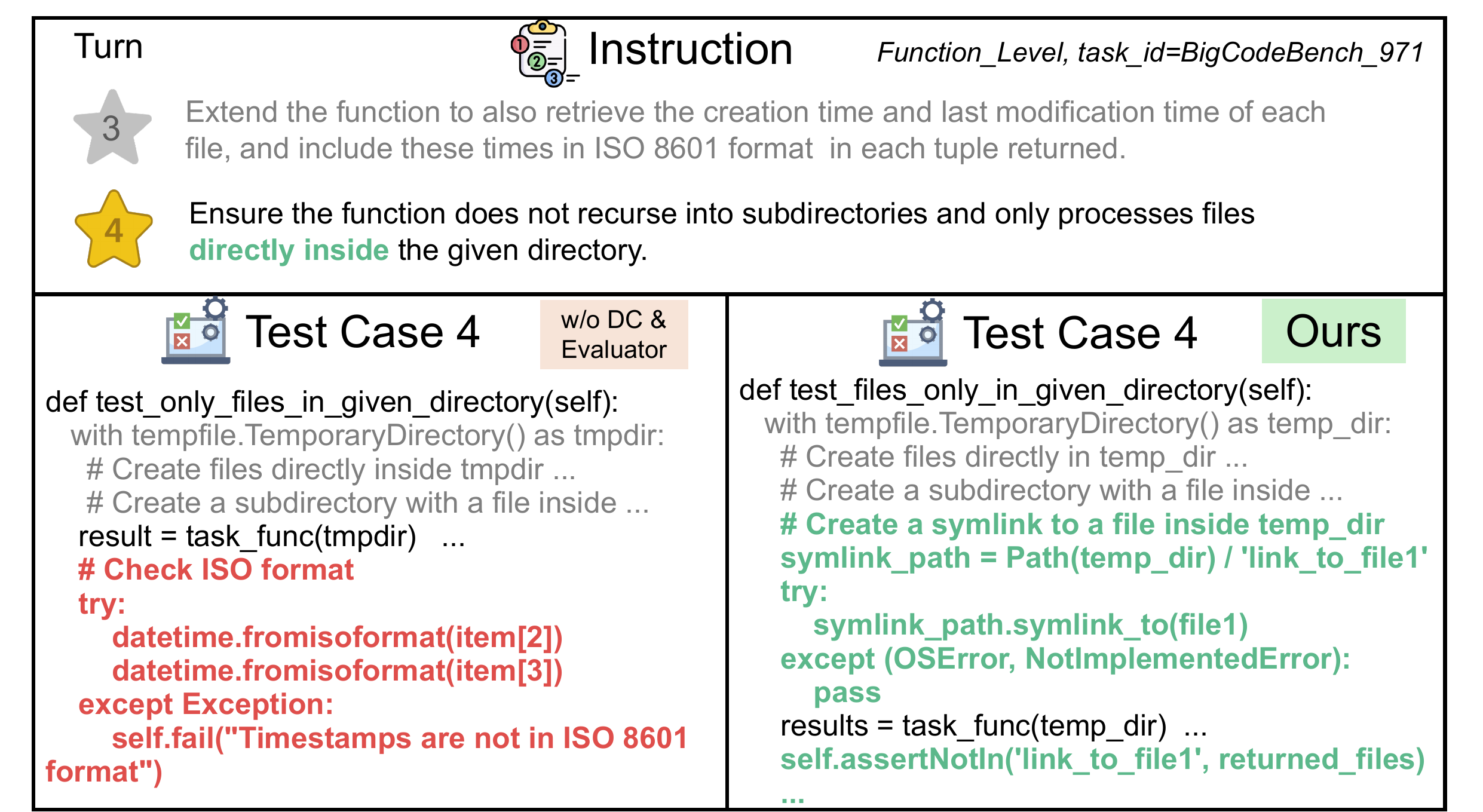} 
    \caption{Case study of test cases generated under two validation strategies.}
    \label{fig:rq3_case_study}
\vspace{-1em}
\end{figure}

As shown in Table~\ref{tab:rq4}, the performance of all LLMs follows a consistent pattern across settings: the lowest accuracy and completion rates are achieved on test suites generated by \textit{Ours}, followed by \textit{w/o Evaluator}, with the highest scores observed in \textit{w/o Distinctiveness Validation \& Evaluator}. 
For instance, on Python tasks, Qwen3-Coder-480B-A35B-Instruct shows a 2.53\% average accuracy and 2.82\% completion rate increase in the \textit{w/o Evaluator} setting versus \textit{Ours}, which surges to a 13.32\% accuracy and 11.97\% completion rate gain in the \textit{w/o Distinctiveness Validation \& Evaluator} setting. This trend holds in Java tasks, with a 0.73\% accuracy and 4.07\% completion rate improvement without the evaluator, climbing to a further 1.93\% in accuracy and 4.65\% in completion rate when both components are disabled. These performance gains underscore that removing either component—especially distinctiveness validation—results in less discriminative and easier-to-pass test cases.

Figure~\ref{fig:rq3_case_study} provides an illustrative example for further illustration.
The task involved a previous turn (Turn 3) where the LLM was instructed to write a function to list files in a directory along with their metadata. In the subsequent turn (Turn 4), the instruction asks the LLM to modify the process method. 
Specifically, the test case generated by our method includes assertions to verify that files within subdirectories were excluded. However, the test case from the \textit{w/o Distinctiveness Validation \& Evaluator} does not include specific checks for the handling of symbolic links at the top level.

\begin{table}[]
\setlength{\tabcolsep}{4.4mm}
\centering
\caption{Performance of LLMs on function-level datasets constructed under different validation strategies. \textbf{Bold} values indicate the lowest scores, reflecting higher dataset distinctiveness within each group. DC indicates Distinctiveness validation stage, and Evaluator indicates LLM-based semantic alignment evaluator. $\Delta_1$ and $\Delta_2$ denote the percentage increase in the average accuracy and complete rate scores, respectively, when compared to our method.}
\label{tab:rq4}
\footnotesize
\scalebox{0.85}{
{
\begin{tabular}{ccccccc}
\toprule
\rowcolor{gray!20} \textbf{Model} &
  \textbf{Language} &
  \textbf{Method} &
  \textbf{Avg Acc$\downarrow$} & 
  \textbf{$\Delta_1$} &
  \textbf{Complete rate$\downarrow$} &
  \textbf{$\Delta_2$}
  \\ \midrule
 &
   &
  \textbf{ours} &
  {\color[HTML]{000000} \textbf{0.4673}} & 0 & 
  {\color[HTML]{000000} \textbf{0.1408}} & 0 \\
 &
   &
  w/o Evaluator &
  0.4740 & \textcolor{red}{0.67\%} &
  0.169 & \textcolor{red}{2.82\%} \\
 &
  \multirow{-3}{*}{\textbf{Python}} &
  w/o DC \& Evaluator &
  0.5848 & \textcolor{red}{11.75\%} &
  0.2183 & \textcolor{red}{7.75\%} \\ \cmidrule{2-7} 
 &
   &
  \textbf{ours} &
  {\color[HTML]{000000} \textbf{0.5254}} & 0 &
  {\color[HTML]{000000} \textbf{0.1977}} & 0\\
 &
   &
  w/o Evaluator &
  0.5289 & \textcolor{red}{0.35\%} &
  0.2035 & \textcolor{red}{0.58\%} \\
\multirow{-6}{*}{DS-T} &
  \multirow{-3}{*}{\textbf{Java}} &
  w/o DC \& Evaluator &
  0.5278 & \textcolor{red}{0.24\%} &
  0.2035 & \textcolor{red}{0.58\%} \\ \midrule
 &
   &
  \textbf{ours} &
  {\color[HTML]{000000} \textbf{0.5225}} & 0 &
  {\color[HTML]{000000} \textbf{0.1972}} & 0 \\
 &
   &
  w/o Evaluator &
  0.5478 & \textcolor{red}{2.53\%} &
  0.2254 & \textcolor{red}{2.82\%} \\
 &
  \multirow{-3}{*}{\textbf{Python}} &
  w/o DC \& Evaluator &
  0.6557 & \textcolor{red}{13.32\%} &
  0.3169 & \textcolor{red}{11.97\%} \\ \cmidrule{2-7} 
 &
   &
  \textbf{ours} &
  {\color[HTML]{000000} \textbf{0.5645}} & 0 &
  {\color[HTML]{000000} \textbf{0.1919}} & 0 \\
 &
   &
  w/o Evaluator &
  0.5718 & \textcolor{red}{0.73\%} &
  0.2326 & \textcolor{red}{4.07\%} \\
\multirow{-6}{*}{QW3-Coder} &
  \multirow{-3}{*}{\textbf{Java}} &
  w/o DC \& Evaluator &
  0.5838 & \textcolor{red}{1.93\%} &
  0.2384 & \textcolor{red}{4.65\%} \\ \midrule
 &
   &
  \textbf{ours} &
  {\color[HTML]{000000} \textbf{0.4915}} & 0 &
  {\color[HTML]{000000} \textbf{0.162}} & 0 \\
 &
   &
  w/o Evaluator &
  0.4984 & \textcolor{red}{0.69\%} &
  0.1972 & \textcolor{red}{3.52\%} \\
 &
  \multirow{-3}{*}{\textbf{Python}} &
  w/o DC \& Evaluator &
  0.5786 & \textcolor{red}{8.71\%} &
  0.2324 & \textcolor{red}{7.04\%} \\ \cmidrule{2-7} 
 &
   &
  \textbf{ours} &
  {\color[HTML]{000000} \textbf{0.5264}} & 0 &
  \textbf{0.1977} & 0 \\
 &
   &
  w/o Evaluator &
  0.5476 & \textcolor{red}{2.12\%} &
  0.2267 & \textcolor{red}{2.9\%} \\
\multirow{-6}{*}{OSS-20-T} &
  \multirow{-3}{*}{\textbf{Java}} &
  w/o DC \& Evaluator &
  0.5426 & \textcolor{red}{1.62\%} &
  0.2035 & \textcolor{red}{0.58\%} \\ \midrule
 &
   &
  \textbf{ours} &
  \textbf{0.4937} & 0 &
 \textbf{0.1901} & 0 \\
 &
   &
  w/o Evaluator &
  0.4945 & \textcolor{red}{0.08\%} &
  \textbf{0.1901} & 0 \\
 &
  \multirow{-3}{*}{\textbf{Python}} &
  w/o DC \& Evaluator &
  0.5404 & \textcolor{red}{4.67\%} &
  0.2183 & \textcolor{red}{4.67\%} \\ \cmidrule{2-7} 
 &
   &
  \textbf{ours} &
  \textbf{0.5308} & 0 &
  {\color[HTML]{000000} \textbf{0.2035}} & 0 \\
 &
   &
  w/o Evaluator &
 0.5347 & \textcolor{red}{0.39\%} &
  0.2209 & \textcolor{red}{1.74\%} \\
\multirow{-6}{*}{OSS-20} &
  \multirow{-3}{*}{\textbf{Java}} &
  w/o DC \& Evaluator &
  0.5403 & \textcolor{red}{0.95\%} &
  0.2093 & \textcolor{red}{0.58\%} \\ \bottomrule
\end{tabular}%
}
}
\vspace{-1em}
\end{table}

\begin{tcolorbox}
\textbf{Finding 4:} {The \textit{semantic-aware discriminative test case generation} method could effectively improve the discriminative ability of test cases in iterative code generation.} 
\end{tcolorbox}

\section{Discussion}\label{sec:discussion}

\subsection{User Study}\label{sec:user_study}

To further investigate how well our multi-turn requirements capture the evolving nature of software specifications in real-world iterative development, we conducted a user study following prior work~\citep{DBLP:conf/icse/GengWD00JML24,DBLP:conf/icse/MuCSWW23,DBLP:journals/pacmse/WuGLWL25}. Specifically, we recruited three developers, each with more than four years of experience in both Python and Java programming.  We randomly sampled 80 tasks from our benchmark (20 from each subset as shown in~\ref{tab:bench_statistics}), resulting in 333 multi-turn interactions. For each task, we presented participants with the requirements, the solution code, and the test cases for every turn in the interaction. Each participant was then asked to provide two sets of ratings on a five-point Likert scale (1 for poor, 2 for marginal, 3 for acceptable, 4 for good, 5 for excellent). An example of questionnaire is shown in our replicate package. First, they evaluated the quality of the code and test cases at each turn. Second, they rated the overall instruction sequence across five dimensions, the details of which are described in Section~\ref{sec:eval_spec}.

The results of our user study are summarized in Table~\ref{tab:user_study}. We observe that both code and test quality at the turn level were rated highly, with average scores above 4.6, while task-level evaluations across all five dimensions also achieved consistently high scores, with averages around 4.7 and low variance. These findings indicate that our constructed multi-turn requirements and corresponding tests are generally perceived as high-quality and well-aligned with real-world iterative development.

\subsection{Implications of Findings}\label{sec:implication}

\subsubsection{Implications for Researchers}

Our findings highlight key limitations of current LLMs in iterative code generation and reveal promising research opportunities.

\begin{itemize}
  \item \textbf{Developing self-correction mechanisms.} The persistent performance gap between the basic and golden settings suggests that LLMs struggle with cumulative context and self-correction. Future work should explore architectures or training methods that incorporate explicit error detection and code-review-like feedback loops to minimize error propagation.
  
  \item \textbf{Refining reasoning strategies.} Since reasoning models often lead to “overthinking” without consistent gains in iterative development, future work could explore dynamically activating reasoning or creating task-aware reasoning modules to better suit iterative development.
  
  \item \textbf{Advancing context management.} The superior performance of lightweight strategies such as \textit{Code Edit} and \textit{Cumulative Instruction} shows the promise of tailored prompt and context management approaches. Techniques like dynamic pruning or compression of irrelevant history may improve both performance and efficiency in long-horizon interactions.

\end{itemize}

\subsubsection{Implications for Developers}

Our findings also offer actionable guidance for practitioners who use LLMs in real-world development workflows.

\begin{itemize}
  \item \textbf{Integrate systematic code review.} When collaborating with LLMs for iterative development, developers should treat model outputs as drafts and apply automated or human-in-the-loop review to prevent error accumulation.
  
  \item \textbf{Cautious use of reasoning-enhanced models.} Models with explicit reasoning do not automatically perform better in iterative settings. Practitioners should validate them under their own multi-turn scenarios or consider lightweight thinking tools as alternatives where appropriate.
  
  \item \textbf{Apply lightweight context strategies.} Instead of feeding full conversation history to the model, use focused strategies such as editing-based prompts or cumulative instructions to retain only task-relevant context. This can reduce token cost and latency while maintaining or improving accuracy.
\end{itemize}


\begin{wrapfigure}{r}{0.5\textwidth} 
    \centering 
    \small
    \captionof{table}{The statistical results of our user study.} 
    \label{tab:user_study}
    \setlength{\tabcolsep}{3pt} 
    \begin{tabular}{c|c|c|c|c}
    \toprule
    \rowcolor{gray!20} \textbf{Type}               & \textbf{Dimension} & \textbf{Count} & \textbf{Avg.} & \textbf{Std.} \\ \midrule
    \multirow{2}{*}{Turn level} & Solution quality   & 999            & 4.650         & 0.518         \\
                                & Test quality       & 999            & 4.745         & 0.463         \\ \midrule
    \multirow{5}{*}{Task level} & Testability        & 240            & 4.808         & 0.405         \\
                                & Completeness       & 240            & 4.729         & 0.464         \\
                                & Distinctiveness    & 240            & 4.712         & 0.472         \\
                                & Authenticity       & 240            & 4.704         & 0.457         \\
                                & Coherence          & 240            & 4.758         & 0.439         \\ \bottomrule
    \end{tabular}%
\vspace{-1em}
\end{wrapfigure}

\subsection{Threats to Validity}\label{sec:discussion_threats_to_validity}

We have identified the following major threats to validity:

\textbf{Limitations in LLM-driven Data Generation.} Our benchmark leverages LLMs for instruction synthesis and test case generation. Due to the inherent hallucination issues in LLMs, this process may generate incorrect instructions and test cases. To mitigate this problem, we propose a multi-agent approach to improve instruction quality and introduce a semantic-aware discriminative test case generation method to generate high-quality test cases.


\textbf{Potential Data Leakage.}  One threat is potential data leakage in our constructed evaluation data. Since all the data used is publicly available, the LLM may have been trained on this data. To mitigate this problem, we select recent popular benchmarks as seed data and leverage a leakage filtering step in seed dataset preparation to remove samples that might have data leakage issue.


\section{Conclusion}\label{conclusion}

In this paper, we present \tool, the first benchmark dedicated to evaluating LLMs for iterative code generation under progressive requirement refinement. The construction of \tool follows a carefully designed pipeline that first leverages a multi-agent-based requirement generation method to generate the multi-round interaction process from final requirements, then employs a semantic-aware discriminative test case generation component to ensure discriminative and consistent evaluation at each turn.  
Our empirical study across 11 state-of-the-art models reveals some findings such as the persistent difficulty of iterative
code generation with evolving requirements, with low completion rates even under ideal conditions. These findings highlight significant gaps between current LLM capabilities and the demands of real-world, iterative programming workflows.

\section*{Data Availability}
Our benchmark and evaluation scripts are publicly available at \url{https://github.com/AjaxZhan/SR-Eval}.


\bibliographystyle{ACM-Reference-Format}
\bibliography{ref.bib}


\end{document}